\documentclass[useAMS,usenatbib]{mn2e}

\date{Accepted 2009 January 12}
\volume{394}
\pagerange{1857--1874}
\pubyear{2009}

\usepackage{txfonts}
\usepackage{amsfonts}
\usepackage{amssymb}
\usepackage{graphicx}
\usepackage{longtable}

\newcommand{\be}{\begin{equation}}
\newcommand{\ee}{\end{equation}}
\newcommand{\bea}{\begin{eqnarray}}
\newcommand{\eea}{\end{eqnarray}}

\newcommand{\f}{R_{\rm mol}}
\newcommand{\fc}{\f^{\rm c}}

\newcommand{\fg}{\f^{\rm galaxy}}
\newcommand{\fu}{\f^{\rm universe}}
\newcommand{\fga}{R_{\rm mol,0}^{\rm galaxy}}
\newcommand{\fgb}{R_{\rm mol,1}^{\rm galaxy}}
\newcommand{\fgc}{R_{\rm mol,2}^{\rm galaxy}}
\newcommand{\fgd}{R_{\rm mol,3}^{\rm galaxy}}
\newcommand{\fgi}{R_{\rm mol,i}^{\rm galaxy}}
\newcommand{\fgj}{R_{\rm mol,j}^{\rm galaxy}}
\newcommand{\fganal}{R_{\rm mol,th}^{\rm galaxy}}
\newcommand{\freq}{\nu}
\newcommand{\h}{h}
\newcommand{\ha}{HI}
\newcommand{\hfraction}{\beta}

\newcommand{\hm}{H$_2$}

\newcommand{\mass}{M}
\newcommand{\magb}{M_{\rm B}}
\newcommand{\mg}{\mass_{\rm gas}}
\newcommand{\mha}{\mass_{\rm HI}}
\newcommand{\mhm}{\mass_{{\rm H}_2}}
\newcommand{\ms}{\mass_{\rm stars}}
\newcommand{\msun}{{\rm M}_{\odot}}
\newcommand{\msdisk}{\mass_{\rm stars}^{\rm disc}}
\newcommand{\Omegag}{\Omega_{\rm gas}}
\newcommand{\Omegaha}{\Omega_{\rm HI}}
\newcommand{\Omegahm}{\Omega_{{\rm H}_2}}

\newcommand{\phiha}{\phi_{\rm HI}}
\newcommand{\phihm}{\phi_{{\rm H}_2}}

\newcommand{\rd}{r_{\rm disc}}

\newcommand{\rhoha}{\rho_{\rm HI}}
\newcommand{\rhohm}{\rho_{\rm H_2}}

\newcommand{\Sigmag}{\Sigma_{\rm gas}}
\newcommand{\Sigmasdisk}{\Sigma_{\rm stars}^{\rm disc}}
\newcommand{\Sigmaha}{\Sigma_{\rm HI}}
\newcommand{\Sigmahm}{\Sigma_{\rm H_2}}
\newcommand{\sigf}{\sigma_{\rm data}}
\newcommand{\sigfi}{\sigma_{\rm data,i}}

\newcommand{\sigo}{\sigma_{\rm obs}}
\newcommand{\sigp}{\sigma_{\rm phy}}
\newcommand{\sigpa}{\sigma_{\rm phy,0}}
\newcommand{\sigpb}{\sigma_{\rm phy,1}}
\newcommand{\sigpc}{\sigma_{\rm phy,2}}
\newcommand{\sigpd}{\sigma_{\rm phy,3}}
\newcommand{\sigpi}{\sigma_{\rm phy,i}}
\newcommand{\sigx}{\sigma_{\rm X}}
\newcommand{\Ss}{\Sigma_{\rm stars}}

\newcommand{\vs}{\sigma_{\rm stars}}
\newcommand{\vsz}{\sigma_{\rm stars,z}}
\newcommand{\vgz}{\sigma_{\rm gas,z}}

\newcommand{\mnras}{MNRAS} %{Mon. Not. R. Astron. Soc.}
\newcommand{\aj}{AJ} %{Astron. J.}
\newcommand{\apj}{ApJ} %{Astrophys. J.}
\newcommand{\apjl}{ApJL} %{Astrophys. J. Lett.}
\newcommand{\aap}{A\&A} %{Astron. Astrophys.}
\newcommand{\apjs}{ApJS} %{Astrophys. J. Suppl.}
\newcommand{\araa}{ARA\&A} %{Annu. Rev. Astron. Astrophys.}
\newcommand{\pasj}{PASJ} %{Pub. Astron. Soc. Japan}
\newcommand{\apss}{Ap\&SS} % Astrophysics and Space Science

\title{Understanding the H$_\mathbf{2}$/HI Ratio in Galaxies}
\author[D. Obreschkow and S. Rawlings]{D. Obreschkow and S. Rawlings\\
Astrophysics, Department of Physics, University of Oxford, Keble Road, Oxford, OX1 3RH, UK}

\begin{document}

\maketitle

\label{firstpage}

\begin{abstract}
We revisit the mass ratio $\fg$ between molecular hydrogen (\hm) and atomic hydrogen (\ha) in different galaxies from a phenomenological and theoretical viewpoint. First, the local \hm-mass function (MF) is estimated from the local CO-luminosity function (LF) of the FCRAO Extragalactic CO-Survey, adopting a variable CO-to-\hm~conversion fitted to nearby observations. This implies an average \hm-density $\Omegahm=(6.9\pm2.7)\cdot10^{-5}\h^{-1}$ and $\Omegahm/\Omegaha=0.26\pm0.11$ in the local Universe. Second, we investigate the correlations between $\fg$ and global galaxy properties in a sample of 245 local galaxies. Based on these correlations we introduce four phenomenological models for $\fg$, which we apply to estimate \hm-masses for each \ha-galaxy in the HIPASS catalog. The resulting \hm-MFs (one for each model for $\fg$) are compared to the reference \hm-MF derived from the CO-LF, thus allowing us to determine the Bayesian evidence of each model and to identify a clear best model, in which, for spiral galaxies, $\fg$ negatively correlates with both galaxy Hubble type and total gas mass. Third, we derive a theoretical model for $\fg$ for regular galaxies based on an expression for their axially symmetric pressure profile dictating the degree of molecularization. This model is quantitatively similar to the best phenomenological one at redshift $z=0$, and hence represents a consistent generalization while providing a physical explanation for the dependence of $\fg$ on global galaxy properties. Applying the best phenomenological model for $\fg$ to the HIPASS sample, we derive the first integral cold gas-MF (\ha+\hm+helium) of the local Universe.
\end{abstract}


\section{Introduction}\label{introduction}

The Interstellar Medium (ISM) plays a vital role in galaxies as their primordial baryonic component and as fuel or exhaust of stars. Hydrogen constitutes 74 per cent of the mass of the ISM. When it is cold and neutral it coexists in the atomic phase (\ha) and molecular phase (\hm). While the former follows a smooth distribution across large galactic substructures, the latter is found in dense molecular clouds \citep{Drapatz1984} acting as the sole cr\`{e}ches of newborn stars. The dissimilar but interlinked roles of \ha~and \hm~in substructure growth and star formation have caused a growing interest in simultaneous observations of both phases and cosmological simulations that distinguish between \ha~and \hm.

Extragalactic observations of \ha~often use its prominent 21-cm emission line, and currently comprise several thousand galaxies \citep[HI Parkes All Sky Survey HIPASS,][]{Barnes2001}, and a maximum redshift of $z=0.2$ \citep{Verheijen2007}. By contrast, most \hm-estimates must rely on indirect tracers, such as CO-lines, with uncertain conversion factors. Consequently, the phase ratio of neutral hydrogen $\f\equiv {\rm d}\mhm/{\rm d}\mha$ and its value for entire galaxies $\fg\equiv\mhm/\mha$ remain debated, and estimates of the universal density ratio $\fu\equiv\Omegahm/\Omegaha$ vary by an order of magnitude in the local Universe (e.g.~0.14, 0.42, 1.1 stated respectively by \citealp{Boselli2002}, \citealp{Keres2003}, \citealp{Fukugita1998}).

Ultimately, the uncertainties of \hm-measurements hinder the reconstruction of cold gas masses $\mg=(\mha+\mhm)/\hfraction$, where
$\hfraction\approx0.74$ is the standard fraction of hydrogen in neutral gas with the rest consisting of helium (He) and a minor fraction of heavier elements. The limitations of comparing $\mha$ to $\mg$ caused by the measurement uncertainties of $\mhm$ culminate in severe difficulties to compare statistically tight cold gas-mass functions (MFs) of modern cosmological simulations with precise HI-MFs extracted from \ha-surveys, such as HIPASS. Both simulations and surveys have reached statistical accuracies far better than any current model for $\fg$, and hence the comparison of observations with simulations is mainly limited by the uncertainty of $\fg$.

\begin{figure}
  \includegraphics[width=\columnwidth]{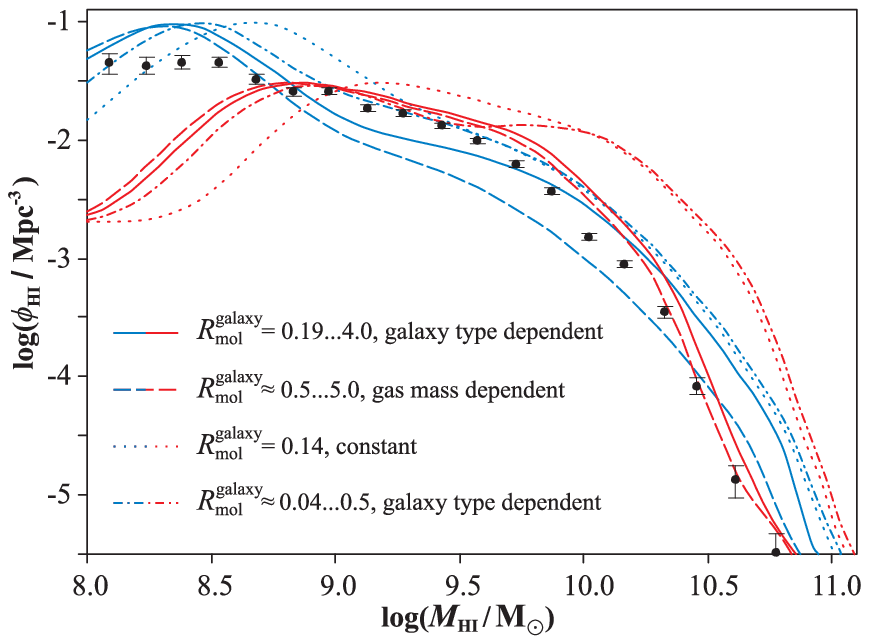}
  \caption{Dots represent the observed HI-MF by \citet{Zwaan2005}; lines represent simulated HI-MFs derived from the semi-analytic models by \citet[][red lines]{Bower2006} and \citet[][blue lines]{DeLucia2007}. The four models of $\fg$ were adopted or derived from \citet[][solid lines]{Young1989b}, \citet[][dashed lines]{Keres2003}, \citet[][dotted lines]{Boselli2002}, and \citet[][dash-dotted lines]{Sauty2003}.}
  \label{fig_uncertain_conversion}
\end{figure}

As an illustration, Fig.~\ref{fig_uncertain_conversion} displays the observed \ha-MF from the HIPASS sample \citep{Zwaan2005} together with several simulated \ha-MFs. The latter are based on the cold gas masses of the simulated galaxies produced by two different galaxy formation models applied to the Millennium Simulation \citep{Bower2006,DeLucia2007}. We have converted these cold gas masses into \ha-masses using four models for $\fg$ from the literature \citep{Young1989b,Keres2003,Boselli2002,Sauty2003}. The figure adopts the Hubble constant of the Millennium Simulation, i.e.~$\h=0.73$, where $h$ is defined by $H_0=100\,\h$\,km\,s$^{-1}$\,Mpc$^{-1}$ with $H_0$ being the present-day Hubble constant. The differential gas density $\phiha$ is defined as $\phiha\equiv{\rm d}\rhoha/{\rm d}\log\mha$, where $\rhoha(\mha)$ is the space density (i.e.~number per volume) of \ha-sources of mass $\mha$. In Fig.~\ref{fig_uncertain_conversion} different models for galaxy formation are distinguished by colour, while the models of $\fg$ are distinguished by line type. Clearly, any conclusion regarding the two galaxy formation models based on their \ha-MFs is affected by the choice of the model for $\fg$.

This paper presents a state-of-the-art analysis of the galaxy-dependent phase ratio $\fg$, the \hm-MF and the integral cold gas-MF (\ha+\hm+He), utilizing various observational constraints. In Section \ref{tracers}, the determination of \hm-masses via CO-lines is revisited and an empirical, galaxy-dependent model for the CO-to-\hm~conversion factor ($X$-factor) is derived from direct measurements of a few nearby galaxies (\citealp{Boselli2002} and references therein). In Section \ref{approach1}, this model is applied to recover an \emph{\hm-MF from the CO-luminosity function (LF)} by \citet{Keres2003}. The resulting \hm-MF significantly differs from the one obtained by \citet{Keres2003} using a constant $X$-factor. Section \ref{approach2} presents an independent derivation of the \emph{\hm-MF from a HI-sample} with well characterized sample completeness \citep[HIPASS,][]{Barnes2001}. This approach is less prone to completeness errors, but it premises an estimate of the \hm/\ha-mass ratio $\fg$. Therefore, we propose four phenomenological models of $\fg$ (as functions of other galaxy properties) and compute their Bayesian evidence by comparing the resulting \hm-MFs to the reference \hm-MF derived from the CO-LF. This empirical method is supported by Section \ref{local}, where we analytically derive a galaxy-dependent model for $\fg$ on the basis of the relation between $\f$ and the pressure of the ISM \citep{Leroy2008}. A brief discussion and a derivation of an integral cold gas-MF (\ha+\hm+He) are presented in Section \ref{discussion}. Section \ref{conclusion} concludes the paper with a summary and outlook.

\section{The variable CO-to-\hm~conversion}\label{tracers}

\subsection{Background: basic mass measurement of \ha~and \hm}\label{background}

\ha~emits rest-frame 1.42 GHz radiation ($\lambda=0.21\ $m) originating from the hyperfine spin-spin relaxation. Especially cold \ha~ \citep[$T\sim50-100\ $K, see][]{Ferriere2001} also appears in absorption against background continuum sources or other \ha-regions, but makes up a negligible fraction in most galaxies. Within this assumption, HI can be considered as optically thin on galactic scales, and hence the HI-line intensity is a proportional mass tracer,
\be
  \frac{\mha}{\msun} = 2.36\cdot10^5\cdot\frac{S_{\rm HI}}{{\rm Jy\,km\,s}^{-1}}\cdot\left(\frac{D_{\rm l}}{{\rm Mpc}}\right)^2,
  \label{eq_mha}
\ee
where $S_{\rm HI}$ is the integrated HI-line flux density and $D_{\rm l}$ is the luminosity distance to the source.

Unlike \ha-detections, direct detections of \hm~in emission rely on weak lines in the infrared and ultraviolet bands \citep{Dalgarno2000} and have so far been limited to the Milky Way and a few nearby galaxies \citep[e.g.][]{Valentijn1999}. Occasionally, \hm~has also been detected at high redshift ($z\approx$2--4) through absorptions lines associated with damped Lyman $\alpha$ systems \citep{Ledoux2003,Noterdaeme2008}. All other \hm-mass estimates use indirect tracers, mostly rotational emission lines of carbon monoxide (CO) -- the second most abundant molecule in the Universe. The most frequently used CO-emission line stems from the relaxation of the $J=1$ rotational state of the predominant isotopomer $^{12}$C$^{16}$O. Radiation from this transition is referred to as CO(1--0)-radiation and has a rest-frame frequency of 115 GHz ($\lambda=2.6\cdot10^{-3}\ $m), detectable with millimeter telescopes. The conversion between CO(1--0)-radiation and \hm-masses is very subtle and generally expressed by the $X$-factor,
\be \label{eqx}
  X \equiv \frac{N_{{\rm H}_2}/{\rm cm}^{-2}}{I_{\rm CO}/({\rm K\,km\,s}^{-1})}\cdot10^{-20},
\ee
where $N_{{\rm H}_2}$ is the column density of molecules and $I_{\rm CO}$ is the integrated CO(1--0)-line intensity per unit surface area defined via the surface brightness temperature $T_{\freq}$ in the Rayleigh-Jeans approximation. Explicitly, $I_{\rm CO}\equiv\int{T_{\freq}{\rm d}V}=\lambda\int{T_{\freq}{\rm d}\freq}$, where $V$ is the radial velocity, $\freq$ is the frequency, and $\lambda=|{\rm d}V/{\rm d}\freq|$ is the wavelength. This definition of the $X$-factor implies a mass-luminosity relation analogous to Eq.~(\ref{eq_mha}) \citep[see review by][]{Young1991},
\be \label{eq_mhm}
  \frac{\mhm}{\msun} = 580\cdot X\cdot\left(\frac{\lambda}{{\rm mm}}\right)^2 \cdot\frac{S_{\rm CO}}{{\rm Jy\,km\,s}^{-1}}\cdot\left(\frac{D_{\rm l}}{{\rm Mpc}}\right)^2,
\ee
where $S_{\rm CO}\equiv\int{S_{{\rm CO,}\freq}{\rm d}V}$ denotes the integrated CO(1--0)-line flux and $D_{\rm l}$ the luminosity distance. $S_{{\rm CO,}\freq}$ is the flux density per unit frequency, for example expressed in Jy, and thus $S_{\rm CO}$ has units like Jy\,km\,s$^{-1}$. Note that $S_{\rm CO}$ relates to the physical flux $F$, defined as power per unit surface, via a factor $\lambda$, i.e.~$F\equiv\int{S_{{\rm CO,}\freq}{\rm d}\freq}=\lambda^{-1}S_{\rm CO}$. CO-luminosities are often defined as $L_{\rm CO}\equiv4\pi D_{\rm l}^2 S_{\rm CO}$ (giving units like Jy\,km\,s$^{-1}$\,(h$^{-1}$\,Mpc)$^2$), thus relating to actual radiative power $P_{\rm CO}$ via $P_{\rm CO}=\lambda^{-1}L_{\rm CO}$. In the $\lambda$-dependent notation above, Eq.~(\ref{eq_mhm}) remains valid for other molecular emission lines, as long as the $X$-factor is redefined with the respective intensities in the denominator of Eq.~(\ref{eqx}).

\subsection{Variation of the $X$-factor among galaxies}\label{bestxfactor}

The theoretical and observational determination of the $X$-factor is a highly intricate task with a long history, and it is perhaps one of the biggest challenges for future CO-surveys.

Theoretically, the difficulty to estimate $X$ arises from the indirect mechanism of CO-emission and from the optical thickness of CO(1--0)-radiation. CO resides inside molecular clouds along with \hm~and acquires rotational excitations from \hm-CO collisions, which can subsequently decay via photon-emission. This mechanism implies that the CO(1--0)-luminosity per unit molecular mass a priori depends on three aspects: (i) the amount of CO per unit \hm, i.e.~the CO/\hm-mass ratio; (ii) the thermodynamic state variables dictating the level populations of CO; (iii) the geometry of the molecular region influencing the degree of self-absorption.

The reason why the CO-luminosity can be used at all as a \hm-mass tracer is a statistical one. In fact, CO-luminosities are normally integrated over kiloparsec or larger scales, such as is inevitable given the spatial resolution of most extragalactic CO-surveys. Therefore, hundreds or thousands of molecular clouds are combined into one measurement, and cloud properties, such as geometries and thermodynamic state variables, probably tend towards a constant average, as long as most lines-of-sight to individual clouds do not pass through other clouds, where they would be affected by self-absorption. The latter assumption seems correct for all but nearly edge-on spiral galaxies \citep{Ferriere2001,Wall2006}. It is hence likely that the different geometries and thermodynamic variables of molecular clouds can be neglected in the variations of $X$ and we expect $X$ to depend most significantly on the average CO/\hm-mass ratio of the considered galaxy or galaxy part. However, the determination of the CO/\hm-ratio is itself difficult and its relation to the overall metallicity of the galaxy is uncertain.

Observational estimations of $X$ require CO-independent \hm-mass measurements, which are limited to the Milky Way and a few nearby galaxies. Typical methods use the virial mass of giant molecular clouds assumed to be completely molecularized \citep{Young1991}, the line ratios of different CO-isotopomers \citep{Wild1992}, mm-radiation from cold dust associated with molecular clouds \citep{Guelin1993}, and diffuse high energy $\gamma$-radiation caused by interactions of cosmic-rays with the ISM \citep{Bertsch1993,Hunter1997}.

Early measurements suggested a fairly constant $X$ in the inner $2-10\ $kpc of the Galaxy, leading several authors to the conclusion that $X$ does not significantly depend on cloud properties and metallicity \citep[e.g.~][]{Young1991}. This finding has recently been supported by \cite{Blitz2007}, who analyzed five galaxies in the local group and found no clear trend between metallicity and $X$. The results of \cite{Young1991} and \cite{Blitz2007} rely on the assumption that molecular clouds are virialized. Using the same method \citet{Arimoto1996} detected strong variations of $X$ amongst galaxies and galactic substructures, and they found the empirical power-law relation $X\propto\rm{(O/H)}^{-1}$. \citet{Israel2000} pointed out that molecular clouds cannot be considered as virialized structures, and using far-infrared measurements rather than the virial theorem, \citet{Israel1997} found an even tighter and steeper relation in a sample of 14 nearby galaxies, $X\propto\rm{(O/H)}^{-2.7}$.

In summary, despite rigorous efforts to measure $X$ and its relation to metallicity, the empirical findings remain uncertain and depend on the method used to measure $X$. Since we cannot overcome this issue, we shall use a model for $X$ that relies on different methods to measure $X$, such as presented by \citet{Boselli2002}. Their sample consists of 14 nearby galaxies covering an order of magnitude in O/H-metalicity. This sample includes early- and late-type spiral galaxies, as well as irregular objects and starbursts. For these galaxies $X$ was determined from three different methods: the virial method, mm-data, and $\gamma$-ray data. Their data varies from $X=0.88$ in the center of the face-on Sbc-spiral galaxy M~51 to $X\approx60$ in NGC~55, a barred irregular galaxy seen edge-on. The high values $(X\gtrsim10)$ are often associated with dwarf galaxies and nearly edge-on spiral galaxies, thus consistent with the interpretation of increased CO(1--0) self-absorption in these objects. Typical values for non-edge-on galaxies lie around $X\approx1-5$.

For the particular data set of \citet{Boselli2002}, we shall check the validity of a constant-$X$ model against variable models for $X$, by comparing their Bayesian evidence -- a powerful tool for model selection \citep[e.g.~][]{Sivia2006}. The underlying idea is that the probability $p(M|d)$ of a model $M$ given the data set $d$ is proportional to the probability $p(d|M)$ of $d$ given $M$, provided the compared models are a priori equally likely (Bayes theorem). The probability $p(d|M)$ is also called the \emph{Bayesian evidence} and can be computed as,
\be\label{equ_pdm}
  p(d|M) = \int_\Omega p(d|\mathbf{\theta},M)\pi(\mathbf{\theta}|M)\rm d\mathbf{\theta}
\ee
where $\mathbf{\theta}$ denotes the vector of free parameters of model $M$ and $\Omega$ the corresponding parameter space; $p(d|\mathbf{\theta},M)$ designates the probability of the data given a parameter choice $\mathbf{\theta}$ and it typically includes measurement uncertainties of the data. The prior knowledge on the parameters is encoded in the probability density function $\pi(\mathbf{\theta}|M)$, which satisfies the normalization condition $\int_\Omega\pi(\mathbf{\theta}|M)\rm d\mathbf{\theta}=1$. Two competing models $M_1$ and $M_2$ are compared by their odds, commonly referred to as the \emph{Bayes factor} $B\equiv p(d|M_1)/p(d|M_2)$. According to Jeffrey's scale \citep{Jeffreys1961} for the strength of evidence, $|\ln B\,|<1$ is \emph{inconclusive}, while $|\ln B\,|=1$ reveals \emph{positive evidence} in favour of model $M_1$ (probability$=$0.750), $|\ln B\,|=2.5$ depicts \emph{moderate evidence} (probability$=$0.923), and $|\ln B\,|=5$ expresses \emph{strong evidence} (probability$=$0.993).

We consider the four models listed in Table \ref{table_xbic}: a constant model, where $\theta=(c_0)$, and three linear models, where $\mathbf{\theta}=(c_0,c_1)$. The data are a sample of 14 nearby galaxies, for which $X$ was measured (Table \ref{tab_x_factor}); $X$-factors and O/H-metallicities are taken from \cite{Boselli2002} and references therein, while $\magb$-magnitudes were taken from the HyperLeda database \citep{Paturel2003}, and CO(1--0)-luminosities $L_{\rm CO}$ were derived from the references indicated in Table \ref{tab_x_factor}.

\begin{table}
\centering
\begin{tabular*}{\columnwidth}{@{\extracolsep{\fill}}lcccc}
\hline \\ [-2ex]
   Model for $\log(X)$ & $c_0$ & $c_1$ & rms & $\ln B$ \\ [0.5ex]
\hline \\ [-2ex]
$c_0$                            & $0.43\pm0.15$  & -               & 0.45 & 0.0 \\
$c_0+c_1\cdot \log(\rm{O/H})$         & $-2.90\pm0.20$ & $-1.02\pm0.05$  & 0.19 & 5.1 \\
$c_0+c_1\cdot (\magb-5\,\log\,\h)$ & $3.67\pm0.25$  & $0.176\pm0.006$ &0.29 & 3.3 \\
$c_0+c_1\cdot \log(L_{\rm CO})$  & $1.85\pm0.15$  & $-0.288\pm0.05$ & 0.29 & 2.5 \\ [0.5ex]
\hline
\end{tabular*}
\caption{Comparison of different models for the $X$-factor: $c_0$ and $c_1$ are the best parameters (Gaussian errors are coupled), rms is the rms-deviation of the data from the model, and $B$ is the Bayes factor of each model with respect to the constant model (first row).}
\label{table_xbic}
\end{table}

\begin{table}
  \centering
\begin{tabular*}{\columnwidth}{@{\extracolsep{\fill}}lcccc}
\hline \\ [-2ex]
  \ Object & $\log(\rm{O/H})^{\rm~(a)}$ & $\magb^{\rm~(b)}$ & $\log(L_{\rm CO})$ & $\log(X)^{\rm~(a)}$ \\
  & & $-5\,\log\,\h$ & & \\ [0.5ex]
\hline \\ [-2ex]
\       SMC &  -3.96  & -16.82 &  -2.04$^{\rm~(c)}$ &  1.00         \\
\   NGC1569 &  -3.81  & -15.94 &  -1.60$^{\rm~(d)}$ &  1.18         \\
\       M31 &  -2.99  & -20.23 &  -1.40$^{\rm~(e)}$ & $0.38\pm0.21$ \\
\      IC10 &  -3.69  & -15.13 &  -1.09$^{\rm~(f)}$ & $0.82\pm0.12$ \\
\       LMC &  -3.63  & -17.63 &  -0.68$^{\rm~(g)}$ &  0.90         \\
\       M81 &  -3     & -19.90 &  -0.07$^{\rm~(h)}$ & -0.15         \\
\       M33 &  -3.22  & -18.61 &  0.20$^{\rm~(i)}$ & $0.70\pm0.11$ \\
\       M82 &  -3     & -17.30 &  0.67$^{\rm~(d)}$ &  0.00         \\
\   NGC4565 &  -      & -21.74 &  1.12$^{\rm~(h)}$ &  0.00         \\
\   NGC6946 &  -2.94  & -20.12 &  1.24$^{\rm~(h)}$ &  0.26         \\
\    NGC891 &  -      & -19.43 &  1.48$^{\rm~(h)}$ &  0.18         \\
\       M51 &  -2.77  & -19.74 &  1.80$^{\rm~(h)}$ & -0.22         \\
\ Milky Way &  -3.1   & -19.63 &               - & $0.19\pm0.01$ \\
\   NGC6822 &  -3.84  & -16.07 &               - & $0.82\pm0.20$ \\ [0.5ex]
   \hline
\end{tabular*}
   \caption{Observational data used for the derivation of a variable $X$-factor (Section \ref{bestxfactor}). $L_{\rm CO}$ is given in units of Jy\,km\,s$^{-1}$\,(h$^{-1}$\,Mpc)$^2$. (a) O/H-metallicities and $X$-factors from \citet{Boselli2002}, (b) absolute, extinction-corrected B-Magnitudes from the HyperLeda database \citep{Paturel2003}, (c) \citet{Rubio1991}, (d) \citet{Young1989a}, (e) \citet{Heyer2000}, (f) \citet{Leroy2006}, (g) \citet{Fukui1999}, (h) \citet{Sage1993}, (i) \citet{Heyer2004}.}
   \label{tab_x_factor}
\end{table}

For practical purposes we limit the parameter space $\Omega$ to $c_0\in[-10,10]$ and $c_1\in[-2,2]$ and take the prior probabilities as homogeneous within $\Omega$, i.e.~$\pi(\mathbf{\theta}|M)=1/|\Omega|$. The probability $p(d|\mathbf{\theta},M)$ in Eq.~(\ref{equ_pdm}) is calculated as the product,
\be
  p(d|\mathbf{\theta},M)=\prod_{\rm i} \frac{1}{\sigma\,\sqrt{2\pi}}\,\exp\left\{\frac{[\log(X^{\rm data}_{\rm i})-\log(X^{\rm model}_{\rm i})]^2}{2\,\sigma^2}\right\}
\ee
where $i$ labels the different galaxies listed in Table \ref{tab_x_factor} and $\sigma$ denotes the measurement uncertainty of $\log(X)$. We set $\sigma$ equal the average value $\sigma=0.13$, for all 14 galaxies. (In fact adopting the specific $\sigma$-values listed in Table $\ref{tab_x_factor}$ leads to very similar results, but could be potentially dangerous as the small value $\sigma=0.01$ of the Milky Way is likely underestimated.)

\begin{figure}
  \includegraphics[width=\columnwidth]{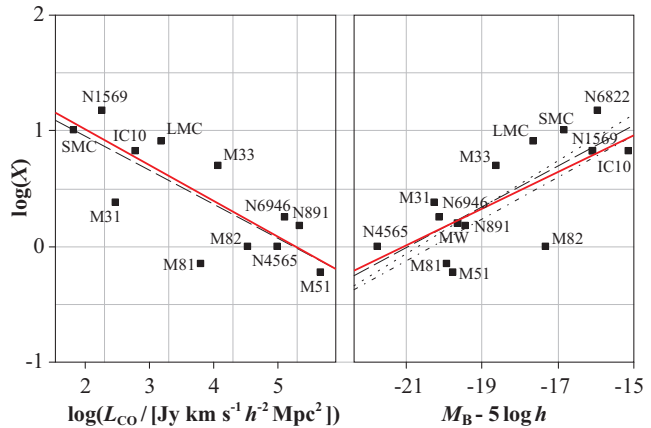}
  \caption{Points represent observed $X$-factors as a function of CO(1--0)-power $L_{\rm CO}$ and absolute blue magnitude $\magb$ for 14 local galaxies. Red solid lines represent linear regressions respecting the mutual relation between $L_{\rm CO}$ and $\magb$ given in Eq.~(\ref{mutualrel}); dashed lines represent independent linear regressions; the dotted line represents the linear fit found by \citet{Arimoto1996}; and the dash-dotted line represents the linear fit found by \citet{Boselli2002}.}\label{fig_x_factor}
\end{figure}

The evidence integrals were solved numerically using a Monte Carlo sampling of the parameter space. The resulting Bayes factors (listed in Table \ref{table_xbic}) reveal moderate to strong Bayesian evidence for a variable $X$-factor given the $X$-factors presented by \cite{Boselli2002}. Among the different variable models for $\log(X)$, the best one depends linearly on $\log(\rm{O/H})$ (highest Bayes factor), as expected from the natural dependence of the CO/\hm~ratio on the O/H ratio. However, $\log(X)$ is also well correlated with $\magb$ and $\log(L_{\rm CO})$, and hereafter we will use those relations because of the widespread availability of $\magb$ and $L_{\rm CO}$ data. In fact, a $X$-factor depending on $L_{\rm CO}$ simply translates to a non-linear conversion of CO-luminosities into \hm-masses. If the two linear regressions between $\log(X)$ and $\magb$ and between $\log(X)$ and $\log(L_{\rm CO})$ were determined independently, they would imply a third linear relation between $\magb$ and $\log(L_{\rm CO})$. The latter can, however, be determined more accurately from larger galaxy samples. The sample presented in Section \ref{observedsample} (245 galaxies) yields
\be\label{mutualrel}
\log(L_{\rm CO})\approx-4.5-0.52\,(\magb-5\,\log\,\h),
\ee
where $L_{\rm CO}$ is taken in units of Jy\,km\,s$^{-1}$\,(h$^{-1}$\,Mpc)$^2$. To get the best result, we imposed this relation, while simultaneously minimizing the square deviations of the two regressions between $\log(X)$ and respectively $\magb$ and $\log(L_{\rm CO})$. In such a way we find
\bea
  \log(X) & = & 1.97-0.308\,\log(L_{\rm CO})\pm\sigx, \label{xp115} \\
  \log(X) & = & 3.36+0.160\,(\magb-5\,\log\,\h)\pm\sigx. \label{xmb}
\eea
These two relations are shown in Fig.~\ref{fig_x_factor} (red solid lines). For comparison, the independent regressions, obtained without imposing the relation given in Eq.~(\ref{mutualrel}), are plotted as dashed lines. These relations correspond to the parameters $c_0$ and $c_1$ given in Table \ref{table_xbic}. Other regressions found by \cite{Arimoto1996} and \cite{Boselli2002} are also displayed. Their approaches are similar, but \cite{Arimoto1996} used less galaxies (8 instead of 14). The 14 data points in Fig.~\ref{fig_x_factor} are scattered around the relations of Eqs.~(\ref{xp115}) and (\ref{xmb}) with the same rms-deviation of $0.29$ in $\log(X)$. Combined with the average measurement uncertainty of $\sigma=0.13$, this gives an estimated true physical scatter in $\log(X)$ of $\sigx=(0.29^2-0.13^2)^{1/2}=0.26$.

The variable models of $X$ given in Eqs.~(\ref{xp115}) and (\ref{xmb}) will be applied in Sections \ref{approach1} and \ref{approach2}. In order to account for the uncertainties of $X$ highlighted in the beginning of this section, we shall also present the results for a constant $X$-factor with random scatter in Section \ref{approach1}.

\section{Deriving the \hm-MF from the CO-LF}\label{approach1}

Using the variable model for the $X$-factor of Eq.~(\ref{xp115}), we shall now recover the local \hm-mass function (\hm-MF) from the CO-LF presented by \citet{Keres2003}. The latter is based on a far infrared-selected subsample of 200 galaxies from the FCRAO Extragalactic CO-Survey \citep{Young1995}, which successfully reproduced the $60\mu$m-LF, thus limiting the errors caused by the incompleteness of the sample. \cite{Keres2003} themselves derived a \hm-MF using a constant model $X=3$, which probably leads to an overestimation of the \hm-abundance, especially in the high mass end, where the $X$-factors tend to be lower according to the data shown in Section \ref{bestxfactor}.

\begin{figure}
  \includegraphics[width=\columnwidth]{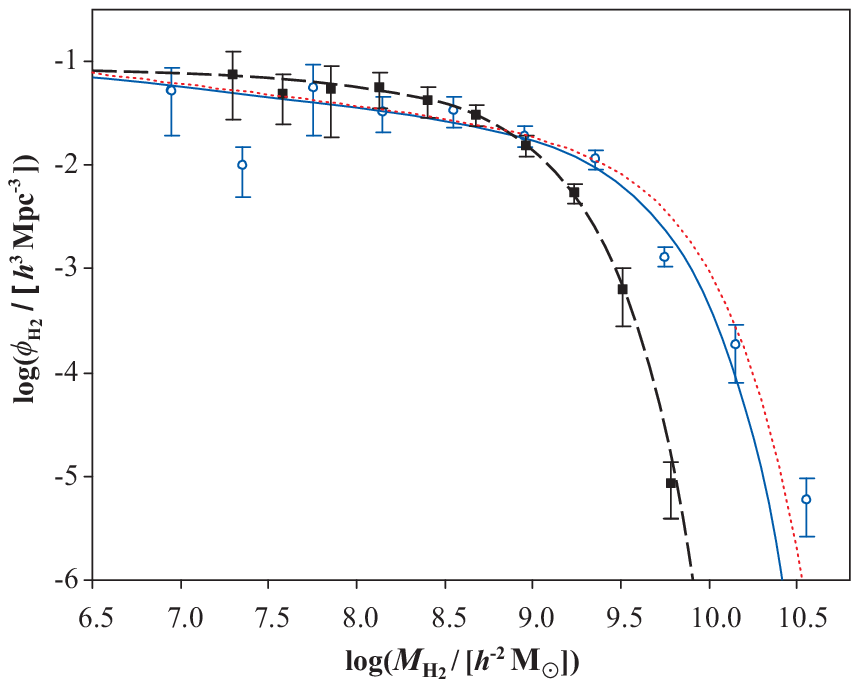}
  \caption{Filled squares represent our reference \hm-MF derived directly from the observed CO-LF \citep{Keres2003} using the variable $X$-factor of Eq.~(\ref{xp115}) with scatter $\sigx=0.26$. Open circles are the original \hm-MF obtained by \citet{Keres2003} using a constant factor $X=3$ without scatter. The dashed and solid lines represent Schechter function fits to our reference \hm-MF and the original \hm-MF, while the dotted line represents the Schechter function corresponding to a constant $X$-factor $X=3$ with scatter $\sigx$.}\label{fig_h2_mf}
\end{figure}

We applied Eq.~(\ref{xp115}) with scatter $\sigx=0.26$ to the individual data points of the CO-LF given by \cite{Keres2003}. The resulting \hm-MF -- hereafter the \emph{reference \hm-MF} -- is shown in Fig.~\ref{fig_h2_mf} together with the \emph{original \hm-MF} derived by \cite{Keres2003} using the constant factor $X=3$ without scatter. To both functions we fitted a Schechter function \citep{Schechter1976} of the form
\be
    \phihm = \ln(10)\cdot\phi^\ast\cdot\left(\frac{\mhm}{\mass^\ast}\right)^{\alpha+1}\exp\left[-\left(\frac{\mhm}{\mass^\ast}\right)\right]
\ee
by minimizing the weighted square deviations of all but the highest \hm-mass bin. \cite{Keres2003} argue that this bin may contain a CO-luminous subpopulation of starburst galaxies, similarly to the situation in the far infrared continuum \citep{Yun2001}. In any case the last bin only marginally contributes to the universal \hm-density. The Schechter function parameters are given in Table \ref{tab_h2mfs}, as well as the reduced $\chi^2$ of the fits, total \hm-densities $\rhohm$ and $\Omegahm\equiv\rhohm/\rho_{\rm crit}$, and the average molecular ratio $\fu\equiv\Omegahm/\Omegaha$. Both $\rhohm$ and $\Omegahm$ were evaluated from the fitted Schechter function rather than the binned data, and $\Omegaha=(2.6\pm0.3)\ \h^{-1}10^{-4}$ was adopted from the HIPASS analysis by \citet{Zwaan2005}.

Our new reference \hm-MF is compressed in the mass-axis compared to the original one, and our estimate of $\rhohm$ (Table \ref{tab_h2mfs}) is 33 per cent smaller. The global \hm/\ha-mass ratio drops to $0.26\pm0.11$, implying a total cold gas density of $\Omegag=(4.4\pm0.8)\cdot10^{-4}\ \h^{-1}$. The composition of cold gas becomes: $59\pm6$ per cent \ha, $15\pm6$ per cent \hm, 26 per cent He and metals, where the uncertainties of \ha~and \hm~are anti-correlated.

It is interesting to observe the quality of the Schechter function fits: the fit to our reference \hm-MF is much better than the one to the original \hm-MF \citep{Keres2003}. Since the original MF is a simple shift of the CO-LF (constant $X$-factor), the Schechter function fit to our reference \hm-MF is also much better than the fit to the CO-LF. We could demonstrate that this difference is partially caused by the scatter $\sigx=0.26$, applied to the variable $X$-factor when deriving the reference \hm-MF from the CO-LF. Scatter averages the densities in neighboring mass bins, hence smoothing the reference MF. Additionally, there is a fundamental reason for the rather poor Schechter function fit of the CO-LF: It is formally impossible to describe both the \hm-MF and the CO-LF with Schechter functions, if the two are interlinked via the linear transformation of Eq.~(\ref{xp115}). Yet, in analogy to the \ha-MF \citep{Zwaan2005}, it is likely that the \hm-MF is well matched by a Schechter function, hence implying that the CO-LF deviates from a Schechter function.

\begin{table}
\centering
\begin{tabular*}{\columnwidth}{@{\extracolsep{\fill}}lcc}
\hline \\ [-2ex]
                              & reference \hm-MF & original \hm-MF \\
                              & (variable $X$) & (constant $X$) \\ [0.5ex]
\hline \\ [-2ex]
\ \ $\mass^\ast$              & $7.5\cdot10^8\,\h^{-2}\,\msun$           &  $2.81\cdot10^9\,\h^{-2}\,\msun$        \\
\ \ $\alpha$                  & $-1.07$                                  &  $-1.18$                                \\
\ \ $\phi^\ast$               & $0.0243\,\h^3\,$Mpc$^{-3}$   &  $0.0089\,\h^3\,$Mpc$^{-3}$ \\
\ \ Red.~$\chi^2$             & $0.05$                                   &  $2.55$                                 \\
\ \ $\rhohm$                  & $(1.9\pm0.7)\cdot10^7\,\h\,\msun\,$Mpc$^{-3}$                      &  $(2.8\pm1.1)\cdot10^7\,\h\,\msun\,$Mpc$^{-3}$                    \\
\ \ $\Omegahm\ $              & $(0.69\pm0.27)\cdot10^{-4}\,\h^{-1} $    &  $(1.02\pm0.39)\cdot10^{-4}\,\h^{-1}$   \\
\ \ $\fu$                     & $0.26\pm0.11$                            &  $0.39\pm0.16$                          \\ [0.5ex]
\hline
\end{tabular*}
\caption{Schechter function parameters, reduced $\chi^2$, and universal mass densities as obtained by integrating the Schechter functions. $\fu\equiv\Omegahm/\Omegaha$ is the global \hm/\ha-mass ratio of the local Universe. The very small reduced $\chi^2$ of our reference \hm-MF arises from a spurious smoothing introduced by the scatter $\sigx$.}
\label{tab_h2mfs}
\end{table}

We finally note, that the faint end of the reference \hm-MF is nearly flat (i.e.~$\alpha=-1$), such that the total \hm-mass is dominated by masses close to the Schechter function break at $\mass^\ast\approx10^9\msun$. In particular, the faint end slope is flatter than for the HI-MF, where $\alpha=-1.37$ \citep{Zwaan2005}, but it should be emphasized that this does not imply that small cold gas masses have a lower molecular fraction. In fact, the contrary is suggested by the observations shown in the Section \ref{approach2}.

For completeness, we re-derived the \hm-MF from the CO-LF using a constant $X$-factor $X=3$ (like \citealp{Keres2003}) with the same Gaussian scatter $\sigx=0.26$ as used for our variable model of $X$. The best Schechter fit for the resulting \hm-MF is also displayed in Fig.~\ref{fig_h2_mf}. The difference between this \hm-MF and the original \hm-MF by \citet{Keres2003} demonstrates that the scatter of $X$ stretches the high mass end towards higher masses.

\section{Phenomenological models for the \hm/\ha-mass ratio}\label{approach2}

In this section, we shall introduce four \emph{phenomenological} models for the \hm/\ha-mass ratio $\fg$ of individual galaxies. Each model will be used to recover a \hm-MF from the HIPASS \ha-catalog \citep{Barnes2001}, thus demonstrating an alternative way to determine the \hm-MF to the CO-based approach. Comparing the \hm-MFs of this section with the reference \hm-MF derived from the CO-LF (Section \ref{approach1}) will allow us to determine the statistical evidence of the models for $\fg$.

\subsection{Observed sample}\label{observedsample}

The sample of galaxies used in this section is presented in Appendix \ref{data} and consists of 245 distinct objects with simultaneous measurements of integrated HI-line fluxes and CO(1--0)-line fluxes. The latter were drawn from 9 catalogs in the literature, and, where not given explicitly, recomputed from indicated \hm-masses by factoring out the different $X$-factors used by the authors. HI line fluxes were taken from HIPASS via the optical cross-match catalog HOPCAT \citep{Doyle2005}. Both line fluxes were homogenized using $\h$-dependent units, where they depend on the Hubble parameter $\h$. Additional galaxy properties were adopted from the homogenous reference database ``HyperLeda'' \citep{Paturel2003}. These properties include numerical Hubble types $T$, extinction corrected blue magnitudes $\magb$, and comoving distances $D_{\rm l}$ corrected for Virgo infall. In the few cases, where these properties were unavailable in the reference catalog, they were copied from the original reference for CO-fluxes. For each galaxy we calculated \ha- and \hm-masses using respectively Eqs.~(\ref{eq_mha}) and (\ref{eq_mhm}). The variable $X$-factors were determined from the blue magnitudes according to Eq.~(\ref{xmb}). We chose to compute $X$ from $\magb$ rather than from $L_{\rm CO}$, because of the smaller measurement uncertainties of the $\magb$ data. Finally, total cold gas masses $\mg=(\mha+\mhm)/\hfraction$ and mass ratios $\fg=\mhm/\mha$ were calculated for each object. While the masses depend on the distances and hence on the Hubble parameter $\h$, the mass ratios $\fg=\mhm/\mha$ are independent of $\h$.

This sample covers a wide range of galaxy Hubble types, masses, and environments, and has 49 per cent overlap with the subsample of the FCRAO Extragalactic CO-Survey used for the derivation of the reference \hm-MF in Section \ref{approach1}. We deliberately limited the sample overlap to 50 per cent in order to control possible sample biases.

We emphasize that this sample exhibits unknown completeness properties, which a priori presents a problem for any empirical model for $\fg$. However, as long as a proposed model is formally complete in the sense that it embodies the essential correlations with a set of free parameters, these parameters can be determined accurately even with an incomplete set of data points. The difficulty in the present case is that no reliable complete model for the molecular fraction $\fg$ has yet been established. We shall bypass this issue by proposing several models for $\fg$ that will be verified with hindsight (Section \ref{fgmodels}). Additional verification will become possible in Section \ref{local}, where we shall derive a physical model for $\fg$.

\subsection{Phenomenological models for $\fg$}\label{fgmodels}

The galaxy sample of Section \ref{observedsample} reveals moderate correlations between $\fg$ and respectively $T$, $\mg$ and $\magb$. These correlations motivate the models proposed below. Other correlations were looked at, such as a correlation between $\fg$ and environment, which may be suspected from stripping mechanisms acting differently on \ha~and \hm. However no conclusive trends could be identified given the observational scatter of $\fg$. All our models are first presented with free parameters, which are fitted to the data at the end of this section.

Model 0 ($\fga$) assumes a constant \hm/\ha-ratio $\fg$, such as is often used in the literature,
\be
  \log(\fga) = q_0+\sigpa, \label{eta0}
\ee
where $q_0$ is a constant and $\sigpa$ denotes an estimate of the physical scatter of perfectly measured data relative to the model.

\begin{figure}
  \includegraphics[width=\columnwidth]{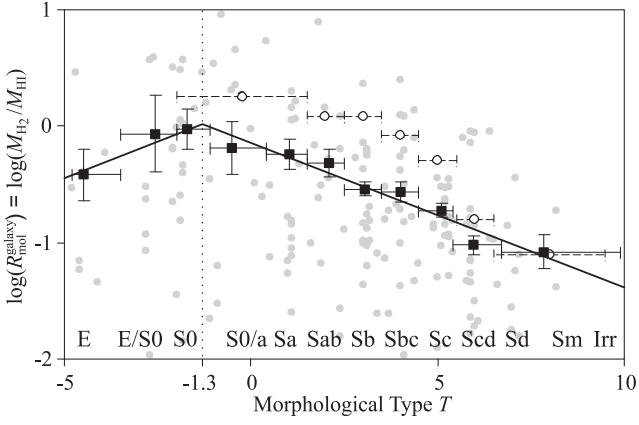}
  \caption{\hm/\ha-mass ratio versus numerical Hubble type $T$. Grey dots represent the empirical data obtained by applying the variable $X$-factor of Eq.~(\ref{xmb}) with scatter to the CO-measurements. Black points represent the binned data; vertical bars represent statistical uncertainties obtained via bootstrapping, i.e.~they depict a 1-$\sigma$ confidence interval of the bin average obtained by examining $10^4$ random half-subsets of the full data; horizontal bars represent the bin intervals. The solid line represents model 1 fitted to the data points. Open circles and dashed bars denote the binned data of the original paper by \citet{Young1989b}.}\label{fig_eta_1}
\end{figure}

Model 1 is galaxy-type dependent, as suggested by earlier studies revealing a trend for $\fg$ to increase from late-type spiral galaxies to early-type ones \citep[e.g.][]{Young1989b,Sauty2003}. The type dependence of our sample is displayed in Fig.~\ref{fig_eta_1}. The binned data clearly show a monotonic increase of the molecular fraction by roughly an order of magnitude when passing from late-type spiral galaxies (Scd--Sd) to early-type spiral and lenticular galaxies (S0--S0/a). The unbinned data illustrate the importance of parameterizing the physical scatter. The Hubble type dependence can be widely explained by the effect of the bulge component on the disc size, as detailed in Section \ref{local}. Observationally, this dependence was first noted by \citet{Young1989b}, whose bins are also displayed in the figure. Their molecular fractions are generally higher, partly due to their rather high assumed $X$-factor of 2.8. The monotonic trend seems to break down between lenticular and elliptical galaxies, where the physical situation becomes more complex. In fact, many elliptical galaxies have molecular gas in their center with no detectable \ha-counterpart, while others seem to have almost no \hm~\citep[e.g.~M 87, see][]{Braine1993}, or may even exhibit \ha-dominated outer regions left over by mergers \citep[e.g.~ NGC 5266, see][]{Morganti1997}. To account for the different behavior of $\fg$ in elliptical and spiral galaxies, we chose a piecewise power-law with different powers for the two populations,
\be\label{eta1}
  \log(\fgb) = \left\{\begin{array}{ll}
                     c^{\rm el}_1+u^{\rm el}_1\,T & {\rm if\ } T<T^\ast_1 \\
                     c^{\rm sp}_1+u^{\rm sp}_1\,T & {\rm if\ } T\geq T^\ast_1
                   \end{array}\right\}+\sigpb
\ee
where $c^{\rm el}_1$, $u^{\rm el}_1$, $c^{\rm sp}_1$, $u^{\rm sp}_1$ are considered as the free parameters to be fitted to the data, and $T^\ast_1$ is at the intersection of the two regressions, i.e.~$c^{\rm el}_1+u^{\rm el}_1\,T^\ast_1\equiv c^{\rm sp}_1+u^{\rm sp}_1\,T^\ast_1$, thus ensuring that $\fgb$ remains a continuous function of $T$ at $T=T^\ast_1$.

\begin{figure}
  \includegraphics[width=\columnwidth]{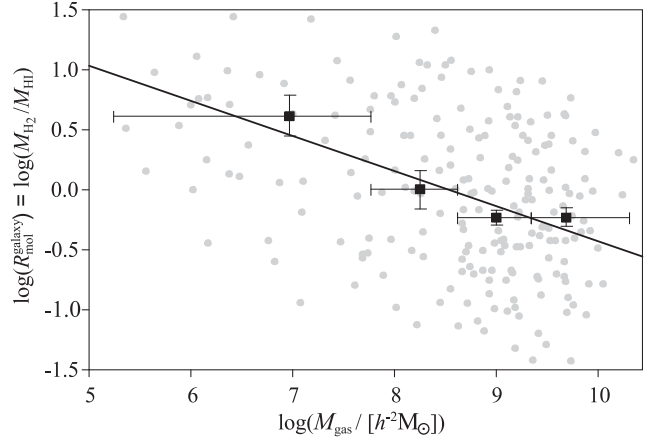}
  \caption{\hm/\ha-mass ratio versus total cold gas mass $\mg\equiv(\mha+\mhm)/\hfraction$. Grey dots represent the empirical data obtained by applying the variable $X$-factor of Eq.~(\ref{xmb}) with scatter to the CO-measurements. Black points represent the binned data; vertical bars represent the 1-$\sigma$ confidence intervals; horizontal bars represent the bin intervals. The solid line represents model 2 fitted to the data points.}\label{fig_eta_2}
\end{figure}

Another correlation exists between $\fg$ and the total cold gas mass $\mg$ or between $\fg$ and the blue magnitude $\magb$. In fact, these two correlations are closely related due to the mutual correlation between $\mg$ and $\magb$, and hence we shall restrict our considerations to the correlation between $\fg$ and $\mg$. According to the roughly monotonic trend visible in Fig.~\ref{fig_eta_2}, we choose a power-law between $\fg$ and $\mg$ for our model 2,
\be
  \log(\fgc) = q_2+k_2\,\log\left(\frac{\mg}{10^9\,\h^{-2}\msun}\right)+\sigpc, \label{eta2}
\ee
where $q_2$, $k_2$ are free parameters. A somewhat similar dependence was recently found between $\fg$ and $\mha$ \citep{Keres2003}, but this result is less conclusive, since $\fg$ and $\mha$ are naturally correlated by the definition of $\fg$, even if $\mha$ and $\mhm$ are completely uncorrelated.

Finally, we shall introduce a fourth model (model 3) for $\fg$ that simultaneously depends on galaxy Hubble type and cold gas mass,
\bea\label{eta3}
  \log(\fgd) &=& \left\{\begin{array}{l}
                     c^{\rm el}_3\,{+u^{\rm el}_3}\,T~~~(\rm{if}~T<T^\ast_3) \\
                     c^{\rm sp}_3{+u^{\rm sp}_3}T~~~(\rm{if}~T\geq T^\ast_3)
                   \end{array}\right\}\\
  &&+k_3\,\log\left(\frac{\mg}{10^9\,\h^{-2}\msun}\right)+\sigpd\,,\nonumber
\eea
where $c^{\rm el}_3$, $u^{\rm el}_3$, $c^{\rm sp}_3$, $u^{\rm sp}_3$, $k_3$ are free parameters and $T^\ast_3$ is defined as $c^{\rm el}_3+u^{\rm el}_3\,T^\ast_3\equiv c^{\rm sp}_3+u^{\rm sp}_3\,T^\ast_3$, thus making $\fgd$ a continuous function of $T$ at $T=T^\ast_3$. Comparing this model with models 1 and 2, will also allow us to study a possible degeneracy between model 1 and model 2 caused by a dependence between cold gas mass and galaxy Hubble type.

The free parameters of the above models were determined by minimizing the rms-deviation between the model predictions and the 245 observed values of $\log(\fg)$ (Appendix \ref{data}). Optimization in log-space is the most sensible choice since $\fg$ is subject to Gaussian scatter in log-space as will be shown in the Section \ref{scatter}. The most probable values of all parameters are shown in Table \ref{table_parameters} together with the corresponding 1-$\sigma$ confidence intervals. The latter were obtained using a bootstrapping method that uses $10^4$ random half-sized subsamples of the full data set and determines the model-parameters for every one of them. The resulting distribution of values for each free parameter was approximated by a Gaussian distribution and its standard deviation $\sigma$ was divided by $\sqrt{2}$ in order to find the 1-$\sigma$ confidence intervals for the full data set. Note that in some cases the parameter uncertainties are coupled, i.e.~a change in one parameter can be accommodated by changing the others, such that the model remains nearly identical. For models 1 and 2, the best fits are displayed in Figs.~\ref{fig_eta_1} and \ref{fig_eta_2} as solid lines.

Table \ref{table_parameters} also shows different scatters that will be explained in Section \ref{scatter}.

\begin{table}
\centering
\begin{tabular*}{\columnwidth}{@{\extracolsep{\fill}}lcccc}
\hline \\ [-2ex]
  Model $\log(\fgi)$ & $i=0$ & $i=1$ & $i=2$ & $i=3$ \\ [0.5ex]
\hline \\ [-2ex]
  $q_{\rm i}$ & $-0.58^{+0.16}_{-0.23}$ & - & $-0.51^{+0.03}_{-0.04}$ & - \\ [0.8ex]
  $c^{\rm el}_{\rm i}$ & - & $+0.18^{+0.40}_{-0.22}$ & - & $-0.01^{+0.25}_{-0.16}$ \\ [0.8ex]
  $u^{\rm el}_{\rm i}$ & - & $+0.12^{+0.14}_{-0.05}$ & - & $+0.13^{+0.07}_{-0.04}$ \\ [0.8ex]
  $c^{\rm sp}_{\rm i}$ & - & $-0.14^{+0.10}_{-0.07}$ & - & $-0.02^{+0.10}_{-0.09}$ \\ [0.8ex]
  $u^{\rm sp}_{\rm i}$ & - & $-0.12^{+0.01}_{-0.02}$ & - & $-0.13^{+0.02}_{-0.02}$ \\ [0.8ex]
  $k_{\rm i}$ & - & - & $-0.24^{+0.05}_{-0.05}$ & $-0.18^{+0.06}_{-0.07}$ \\ [0.8ex]
  $T^\ast_{\rm i}$ & - & $-1.3^{+1.2}_{-0.5}$ & - & $-0.1^{+1.2}_{-0.6}$ \\ [0.5ex]
\hline \\ [-2ex]
  $\sigfi$ & $0.71$ & $0.66$ & $0.67$ & $0.62$ \\
  $\sigpi$ & $0.39$ & $0.27$ & $0.30$ & $0.15$ \\ [0.5ex]
\hline
\end{tabular*}
\caption{The upper panel lists the most likely parameters and 1-$\sigma$ confidence intervals of the four models $\fgi$ ($i=0,...,3$). The bottom panel shows the rms-deviations $\sigfi$ of the data from the model predictions and the estimated physical scatter $\sigpi$ for each model $i$.}
\label{table_parameters}
\end{table}

\subsection{Scatter and uncertainty}\label{scatter}
\begin{figure}
  \includegraphics[width=\columnwidth]{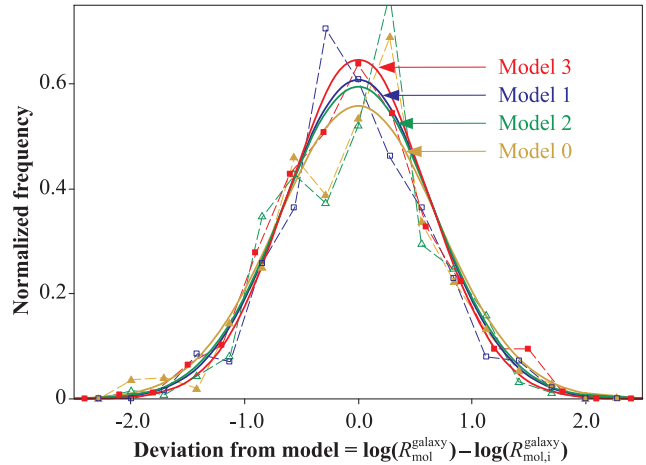}
  \caption{Distributions of the deviations between the observed values of $\log(\fg)$ and the model-values $\log(\fgi)$ ($i=1,...,3$). Data points and dashed lines represent the actual distribution of the data; solid lines represent Gaussian distributions with equal standard deviations.}\label{fig_scatter}
\end{figure}

The empirical values of $\fg$ scatter around the model predictions according to the distributions shown in Fig.~\ref{fig_scatter} (dashed lines). The close similarity of these distributions to Gaussian distributions in log-space (solid lines) allows us to consider the rms-deviations of the data $\sigf$ as the standard deviations of Gaussian distributions. This exhibits the advantage that $\sigf$ can be decomposed in model-independent observational scatter $\sigo$ and model-dependent physical scatter $\sigp$ via the square-sum relation $\sigfi^2=\sigo^2+\sigpi^2$, $i=0,...,3$.

The major contribution to $\sigfi$ comes from observational scatter, as suggested by the close similarity of the different values of $\sigfi$. Indeed, the observational scatter inferred from the $\fg$ values of the 22 repeated sources in our data is $\sigo\approx0.6$. This scatter is a combination of CO-flux measurement uncertainties, uncertain CO/\hm-conversions and \ha-flux uncertainties (in decreasing significance). Since $\sigo$ is only marginally smaller than $\sigfi$ for all models, the estimation of the physical scatters $\sigpi$ (given in Table \ref{table_parameters}) is uncertain. Nevertheless, we shall include these best guesses of the physical scatter, when constructing the \hm-MFs in Section \ref{modelevidence}.

\subsection{Recovering the \hm-MF and model evidence}\label{modelevidence}

Given a model for $\fg$, \hm-masses of arbitrary \ha-galaxies can be estimated. We shall apply this technique to the 4315 sources in the HIPASS catalog using our four models of $\fgi$, $i=0,...,3$. For each model, the resulting \hm-catalog with 4315 objects will be converted into a \hm-MF, which can be compared to our reference \hm-MF derived directly from the CO-LF (Section \ref{approach1}).

For the models $\fgb(T)$ and $\fgd(\mg,T)$ Hubble types $T$ were drawn from the HyperLeda database for each galaxy in the HIPASS catalog by means of the galaxy identifiers given in the optical cross-match catalog HOPCAT \citep{Doyle2005}. \hm-masses were then computed via $\mhm=\fgi\,\mha$, $i=0,...,3$. This equation is implicit in case of the mass-dependent models $\fgc(\mg)$ and $\fgd(\mg,T)$, where $\mg=(\mha+\mhm)/\hfraction$. All four models were applied with scatter, randomly drawn from a Gaussian distribution with the model-specific scatter $\sigpi$, listed in Table \ref{table_parameters}.

\begin{figure}
  \includegraphics[width=\columnwidth]{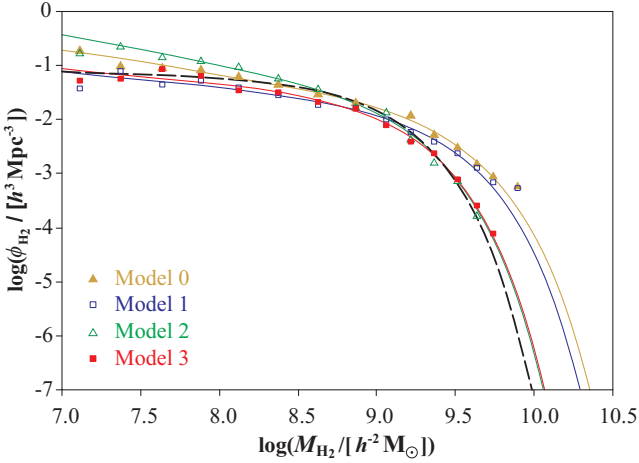}
  \caption{\hm-MFs constructed from the HIPASS \ha-catalog using the different phenomenological models for the \ha/\hm~ratio. The black dashed line is the reference \hm-MF derived from the CO-LF in Section \ref{approach1}.}\label{fig_model_mf}
\end{figure}

In order to reconstruct a \hm-MF for each model, we employed the $1/V_{\rm max}$ method \citep{Schmidt1968}, where $V_{\rm max}$ was calculated from the analytic completeness function for HIPASS that depends on the HI peak flux density $S_{\rm p}$, the integrated HI line flux $S_{\rm int}$, and the flux limit of the survey \citep{Zwaan2004}. After ensuring that we can accurately reproduce the \ha-MF derived by \citet{Zwaan2005}, we evaluated the four \hm-MFs (one for each model $\fgi$) displayed in Fig.~\ref{fig_model_mf} (dots). The uncertainties of $\log(\phi_{\rm H_2})$ vary around $\sigma=0.03-0.1$. Each function was fitted by a Schechter function by minimizing the weighted rms-deviation (colored solid lines).

The comparison of these four \hm-MFs with the reference \hm-MF derived from the CO-LF allows us to qualify the different models $\fgi$, $i=0,...,3$, against each other. We ask: ``What are the odds of model $\fgi$ against model $\fgj$ if the reference \hm-MF derived from the CO-LF is correct?'' This question takes us back to the Bayesian framework of model selection applied in Section \ref{bestxfactor}: If the models are a priori equally likely, their odds are equal to the Bayes factor, defined as the ratio between the model evidences. When computing these evidences, we take the ``observational'' data $d$ to be the reference \hm-MF (with scatter), while the ``model'' data is the \hm-MF reproduced by applying a model $\fgi$ to the HIPASS data. The free parameters $\mathbf{\theta}$ (vector) are listed in Table \ref{table_parameters} for each model (e.g.~$c^{\rm el}_1$, $u^{\rm el}_1$, $c^{\rm sp}_1$, $u^{\rm sp}_1$ for model $\fgb$). The prior probability density $\pi(\mathbf{\theta}|M_{\rm i})$ in the evidence integral of Eq.~(\ref{equ_pdm}) is taken as the multi-dimensional parameter probability distribution function obtained from the 245 galaxies studied in Section \ref{fgmodels} (see Table \ref{table_parameters}). The second piece in the evidence integral, i.e.~the probability density $p(d|\mathbf{\theta},M_{\rm i})$, is calculated as the product,
\be\label{pdm_prod}
  p(d|\mathbf{\theta},M_{\rm i})=\prod_{\rm k} \frac{1}{\sigma\,\sqrt{2\pi}}\,\exp\left[\frac{(\phi^{\rm ref}_{\rm k}-\phi^{\rm model,i}_{\rm k})^2}{2\,\sigma^2}\right]
\ee
where $k$ labels the different bins of the \hm-MF (Fig.~\ref{fig_model_mf}), and $\phi^{\rm ref}_{\rm k}$ and $\phi^{\rm model,i}_{\rm k}$ respectively denote the differential mass densities of the reference \hm-MF and the \hm-MFs reconstructed from HIPASS using the models $\fgi$, $i=0,...,3$. $\sigma$ denotes the combined statistical uncertainties of $\phi^{\rm ref}_{\rm k}$ and $\phi^{\rm model,i}_{\rm k}$, theoretically given by $\sigma^2={\sigma^{\rm ref}_{\rm k}}^2+{\sigma^{\rm model,i}_{\rm k}}^2$. However, we shall neglect the contribution of $\sigma^{\rm model,i}_{\rm k}$, since $\sigma^{\rm ref}_{\rm k}$ is about $3-4$ times larger due to the small size of the FCRAO sample of CO-galaxies compared to the HIPASS sample of HI-galaxies. Furthermore, we assume that $\sigma$ is independent of the bin $k$ and adopt an average uncertainty equal to $\sigma=0.15$ dex. This is the mean scatter of the binned data of the reference \hm-MF (see Fig.~\ref{fig_h2_mf}). Assuming a constant scatter for the whole reference MF artificially increases the weight of the low and high mass ends, where the scatter is indeed closer to 0.3 dex, and reduces the weight of the central part, where the scatter equals 0.1 dex. We argue that this is a reasonable choice, since the central part of the reference \hm-MF suffers most from systematical uncertainties of the $X$-factor and the low and high mass ends encode much of the physics that could discriminate our models for $\fg$ against each other. In any case, the outcome of this evidence analysis is only weakly affected by the choice of scatter.

The integration of the evidence integral is computationally expensive: for \emph{each} choice of model-parameters the following three steps need to be performed: (i) evaluation of the \hm-masses for each galaxies in the HIPASS sample, (ii) computation of the \hm-MF from that sample, (iii) computation of the product in Eq.~(\ref{pdm_prod}). We applied a Monte Carlo method to sample the parameter spaces of the different models. About $10^6$ integration steps had to be performed in total to reach a 2 per cent convergence of the Bayes factors.

\begin{table}
\centering
\begin{tabular*}{\columnwidth}{@{\extracolsep{\fill}}lcccc}
\hline \\ [-2ex]
  Model & $\fga$ & $\fgb$ & $\fgc$ & $\fgd$ \\ [0.5ex]
\hline \\ [-2ex]
  Nb.~of free param. & 1 & 4 & 2 & 5 \\
  $\ln B$ & 0.0 & 7.3 & 8.2 & 22 \\ [0.5ex]
\hline
\end{tabular*}
\caption{Comparison of different models for \hm/\ha-mass ratios of entire galaxies; the first row shows the number of free parameters, while the second row shows the model evidence in terms of the Bayes factor between that model and the constant model $\fga$.}
\label{table_bayesfactors}
\end{table}

The Bayes factor between each model $\fgi$, $i=0,...,3$, and $\fga$ is shown in Table \ref{table_bayesfactors}: We find strong evidence for all variable models ($\fgb$, $\fgc$, $\fgd$) against the constant one ($\fga$), and there is even stronger evidence of the bilinear model ($\fgd$) against all others. The \hm-MF associated with this model is indeed the only one providing a simultaneous fit to the low and high mass ends of the reference MF (see Fig.~\ref{fig_model_mf}), and the agreement is very good (reduced $\chi^2=0.8$).

On a physical level, there are good reasons for the partial failure of the other models in reproducing the extremities of the reference \hm-MF. Model $\fgb(T)$ overestimates the space density of galaxies with high \hm-masses by overestimating $\fg$ for the gas-richest early-type spiral galaxies. In reality, the latter have a very low molecular fraction (see data, model $\fgc$, theory in Section~\ref{local}), but they are a minority within otherwise gas-poor but molecule-rich early-type spirals. Hence, a model depending on Hubble type alone is likely to miss out such objects, resulting in an increased density of high \hm-masses. While model $\fgc(\mg)$ overcomes this issue and produces the right density of high \hm-masses, it fails by a factor $3-4$ in the low-mass end ($\mhm\lesssim10^8\msun$). This is a direct manifestation of assigning high molecular fractions to all gas-poor galaxies, which neglects small young spirals with a dominant atomic phase. Finally, model $\fga$ seems to suffer from limitations at both ends of the \hm-MF.

The clear statistical evidence for model 3 shall be supported by the theoretical derivation of $\fg$ presented in Section \ref{local}.

\section{Theoretical model for the \hm/\ha-mass ratio}\label{local}

So far, we have approached the galactic \hm/\ha-mass ratios $\fg$ with a set of \emph{phenomenological} models, limited to the local Universe. By contrast, we have recently derived a \emph{physical} model for the \hm/\ha-ratios in regular galaxies, which potentially extends to high redshift \citep{Obreschkow2009b}. This model relies on the theoretically and empirically established relation between interstellar gas pressure and local molecular fraction \citep{Elmegreen1993,Blitz2006,Krumholz2009,Leroy2008}. In this section, we will show that the physical model predicts \hm/\ha-ratios consistent with our phenomenological model 3 given in Eq.~(\ref{eta3}). Hence, the physical (or ``theoretical'') model provides a reliable explanation for the global phenomenology of the \hm/\ha-ratio in galaxies.

\subsection{Background: the $\f$--pressure relation}\label{fpbackground}

Understanding the observed continuous variation of $\f$ within individual galaxies \citep[e.g.][]{Leroy2008} requires some explanation, since, fundamentally, there is no mixed thermodynamic equilibrium of \ha~and \hm. To first order, the ISM outside molecular clouds is atomic, while a cloud-region in local thermodynamic equilibrium (LTE) is either fully atomic or fully molecular, depending on the local state variables. The apparent continuous variation of $\f$ is the combined result of (i) a non-resolved conglomeration of fully atomic and fully molecular clouds, (ii) clouds with molecular cores and atomic shells in different LTE, and (iii) some cloud regions off LTE with actual transient mixtures of \ha~and \hm. However, a time-dependent model for off-equilibrium clouds \citep{Goldsmith2007} revealed that the characteristic time taken between the onset of cloud compression and full molecularization is of the order of $10^7$ yrs, much smaller than the typical age of molecular clouds, and hence the fraction of these clouds is small. Therefore, averaged over galactic parts (hundreds or thousands of clouds), $\f$ is dictated by clouds in LTE, entirely defined by a number of state variables.

A theoretical frame exploiting the LTE of molecular clouds was introduced by \citet{Elmegreen1993}, who considered an idealized double population of homogeneous diffuse clouds and isothermal self-gravitating clouds, both of which can have atomic and molecular shells. In this model the molecular mass fraction $f_{\rm mol}={\rm d}\mhm/{\rm d}(\mha+\mhm)$ of each cloud depends on the density profile and the photodissociative radiation density from stars $j$, corrected for self-shielding by the considered cloud, mutual shielding among different clouds, and dust extinction. Since the shielding from this radiation depends on the gas pressure, \cite{Elmegreen1993} finds that $f_{\rm mol}$ essentially scales with the external pressure $P$ and photodissociative radiation density $j$, approximately following $f_{\rm mol}\propto P^{2.2}\,j^{-1}$ with an asymptotic flattening towards $f_{\rm mol}=1$ at high $P$ and low $j$. This implies approximately $\f\equiv {\rm d}\mhm/{\rm d}\mha\propto P^{2.2}\,j^{-1}$. Assuming that $j$ is proportional to the surface density of stars $\Ss$ and that the stellar velocity dispersion $\vs$ varies radially as $\Ss^{0.5}$, \citet{Wong2002} and \citet{Blitz2004,Blitz2006} find roughly $j\propto P$ and hence $\f\propto P^{\,\alpha}$ with $\alpha=1.2$. Recently, \cite{Krumholz2009} have presented a more elaborate theory concluding that $\alpha\approx0.8$. However, the exponent $\alpha$ remains uncertain, thus requiring an empirical determination.

Observationally, \citet{Blitz2004,Blitz2006} were the first ones to reveal a surprisingly tight power-law relation between pressure and molecular fraction based on a sample of 14 nearby galaxies including dwarf galaxies, \ha-rich galaxies, and \hm-rich galaxies. Perhaps the richest observational study published so far is the one by \citet{Leroy2008}, who analyzed 23 galaxies of The HI Nearby Galaxy Survey (THINGS, \citealp{Walter2008}), for which \hm-densities had been derived from CO-data and star formation densities. This analysis confirmed the power-law relation
\be\label{fprelation}
  \f = (P/P_\ast)^\alpha,
\ee
where $P$ is the local, kinematic midplane pressure of the gas, and $P_\ast$ and $\alpha$ are free parameters, whose best fit to the data is given by $P_\ast=2.35\cdot10^{-13}\ {\rm Pa}$ and $\alpha=0.8$.

\subsection{Physical model for the \hm/\ha-ratio in galaxies}\label{etamodel}

We shall now consider the consequence of the model given in Eq.~(\ref{fprelation}) for the \hm/\ha-ratio of entire galaxies. To this end, we adopt the models and methods presented in \cite{Obreschkow2009b}, restricting this paragraph to an overview.

First, we note that most cold gas of regular galaxies is normally contained in a disc. This even applies to bulge-dominated early-type galaxies, such as suggested by recently presented CO-maps of five nearby elliptical galaxies \citep{Young2002}. Hence, the \ha- and \hm-distributions of all regular galaxies can be well described by surface density profiles $\Sigmaha(r)$ and $\Sigmahm(r)$.
We assume that the disc is composed of axially symmetric, thin layers of stars and gas, which follow an exponential density profile with a generic scale length $\rd$, i.e.
\be\label{eqsigma}
  \Sigmasdisk(r)\sim\Sigmag(r)\sim\Sigmaha(r)+\Sigmahm(r)\sim\exp(-r/\rd),
\ee
where $r$ is the galactocentric radius and $\Sigma$ denotes the mass column densities of the different components. Next, we adopt the phenomenological relation of Eq.~(\ref{fprelation}), i.e.~
\be\label{fprelation2}
  \frac{\Sigmaha(r)}{\Sigmahm(r)} = [P(r)/P_\ast]^\alpha,
\ee
and substitute the kinematic midplane pressure $P(r)$ for \citep{Elmegreen1989}
\be\label{eqpr}
  P(r) = \frac{\pi}{2}\,G\,\Sigmag(r)\Big(\Sigmag(r)+f\,\Sigmasdisk(r)\Big),
\ee
where $G$ is the gravitational constant and $f\equiv\vgz/\vsz$ is the ratio between the vertical velocity dispersions of gas and stars. We adopt $f=0.4$ according to \cite{Elmegreen1989}.

Eqs.~(\ref{eqsigma}, \ref{fprelation2}) can be solved for $\Sigmaha(r)$ and $\Sigmahm(r)$. In \cite{Obreschkow2009b}, we demonstrate that the resulting surface profiles are consistent with the empirical data of the two nearby spiral galaxies NGC 5055 and NGC 5194 \citep{Leroy2008}. Integrating $\Sigmaha(r)$ and $\Sigmahm(r)$ over the exponential disc gives the gas masses $\mha$ and $\mhm$, hence providing an estimate of their ratio $\fg$. Analytically, $\fg$ is given by an intricate expression, which is well approximated (relative error $<0.05$ for all galaxies) by the double power-law
\be\label{eqfganal}
  \fganal = \big(3.44\,{\fc}^{-0.506}+4.82\,{\fc}^{-1.054}\big)^{-1},
\ee
where
\be\label{eqfc}
  \fc = \left[11.3\rm\,m^4\,kg^{-2}\,\rd^{-4}\,\mg\,\big(\mg+0.4\,\msdisk\big)\right]^{0.8}.
\ee
$\fc$ is a dimensionless parameter, which can be interpreted as the \hm/\ha-ratio at the center of a pure disc galaxy. For typical cold gas masses of average galaxies ($\mg=10^8-10^{10}\msun$) and corresponding stellar masses and scale radii, $\fc$ calculated from Eq.~(\ref{eqfc}) varies roughly between 0.1 and 50. Hence, $\fg$ given in Eq.~(\ref{eqfganal}) varies roughly between 0.01 and 1.

In summary, Eqs.~(\ref{eqfganal}, \ref{eqfc}) represent a theoretical model for $\fg$, which uses three input parameters: the disc stellar mass $\msdisk$, the cold gas mass $\mg$, and the exponential scale radius $\rd$ (see \citealp{Obreschkow2009b} for a detailed discussion).

\subsection{Mapping between theoretical and phenomenological models}\label{mapping}

We shall now show that our theoretical model for galactic \hm/\ha-mass ratios given in Eqs.~(\ref{eqfganal}, \ref{eqfc}) closely matches the best phenomenological model given in Eq.~(\ref{eta3}). The mapping between the two models uses a list of empirical relations derived from observations of nearby spiral galaxies, and hence the comparison of the models is a priori restricted to spiral galaxies in the local Universe.

\begin{figure}
  \includegraphics[width=\columnwidth]{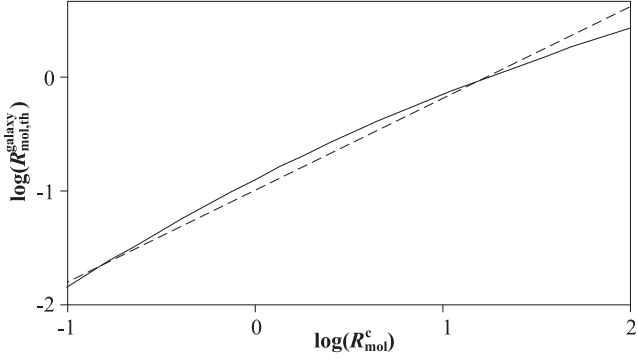}
  \caption{Visualization of the function $\fganal(\fc)$. The solid line represents the nearly exact function given in Eq.~(\ref{eqfganal}), while the dashed line is the power-law fit of Eq.~(\ref{fgapprox}).} \label{fig_fdapprox}
\end{figure}

First, we note that Eq.~(\ref{eqfganal}) can be well approximated by the power-law
\be\label{fgapprox}
  \fganal\approx0.1\,{\fc}^{\,0.8}.
\ee
As shown in Fig.~\ref{fig_fdapprox}, this approximation is accurate to about 10 per cent over the whole range $\fc=0.1,...,50$, covering most regular galaxies in the local Universe.

Substituting $\fc$ in Eq.~(\ref{fgapprox}) for Eq.~(\ref{eqfc}), yields the approximate relation
\be\label{fgapprox2}
  \fganal = \left[0.31\rm\,m^4\,kg^{-2}\,\rd^{-4}\,\mg\,\big(\mg+0.4\,\msdisk\big)\right]^{0.64}.
\ee

In order to compare the theoretical model of $\fg$ to the empirical models of Section \ref{fgmodels}, we need to eliminate the formal dependence of $\fganal$ on $\rd$ and $\msdisk$. To this end, we use two approximate empirical relations, derived from samples of nearby spiral galaxies (see Appendix \ref{obsrelations}),
\bea
  \log\left(\frac{\msdisk}{\h^{-2}\msun}\right)   & = & \gamma_1+\alpha_1\,\log\left(\frac{\mg}{2\cdot10^9\,\h^{-2}\,\msun}\right), \label{phenrel1} \\
  \log\left(\frac{\rd}{\h^{-1}{\rm kpc}}\right) & = & \gamma_2+\alpha_2\,\log\left(\frac{\msdisk}{5\cdot10^9\,\h^{-2}\,\msun}\right)+\delta\,\tilde{T}, \label{phenrel2}
\eea
where $\tilde{T}\equiv(10-T)/16$ is the normalized Hubble type, which varies between $\tilde{T}=0$ (pure disc galaxies) to $\tilde{T}=1$ (pure spheroids).

The parameters corresponding to the best $\chi^2$ fit (Appendix \ref{obsrelations}) are $\alpha_1=1.46\pm0.1$, $\gamma_1=9.80\pm0.05$, $\alpha_2=0.45\pm0.05$, $\gamma_2=0.97\pm0.05$, $\delta=-1.07\pm0.1$. The given intervals are the 1-$\sigma$ confidence intervals of the parameters; they do not characterize the scatter of the data. The units on the right hand side of Eqs.~(\ref{phenrel1}, \ref{phenrel2}) were chosen such as to minimize the correlations between the uncertainties of $\alpha_{\rm i}$ and $\gamma_{\rm i}$.

\begin{figure}
  \includegraphics[width=\columnwidth]{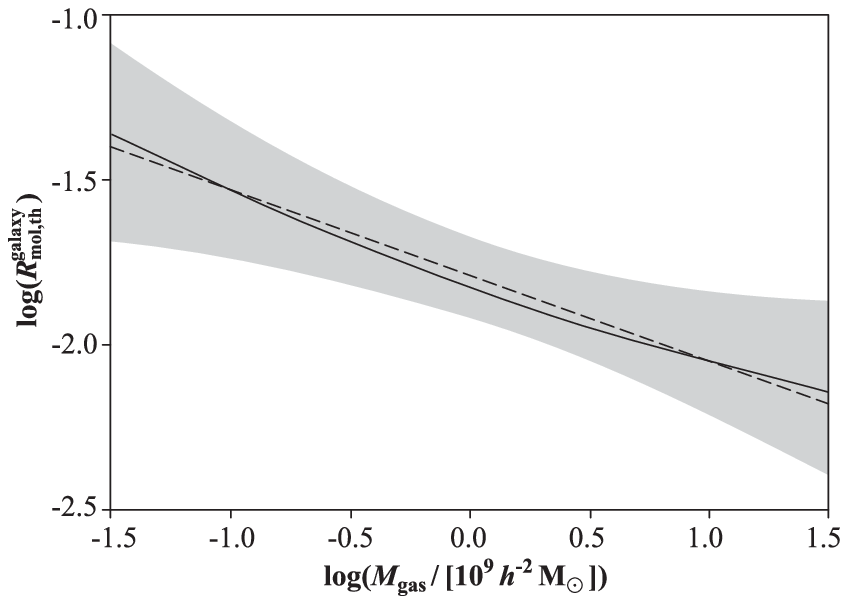}
  \caption{Relation between $\fganal$ and $\mg$ for flat disks ($T=10$). The solid line represents the relation obtained from  Eq.~(\ref{fgapprox2}), when expressing $\rd$ and $\msdisk$ as functions of $\mg$ using Eqs.~(\ref{phenrel1}, \ref{phenrel2}). The shaded zone represents the 1-$\sigma$ uncertainty implied by the uncertainties of the empirical parameters in Eqs.~(\ref{phenrel1}, \ref{phenrel2}). The dashed line represents the best power-law fit for the displayed mass interval as given in Eq.~(\ref{eqfganal3}).}\label{fig_etadiscmg}
\end{figure}

Physical reasons for the empirical relations in Eqs.~(\ref{phenrel1}, \ref{phenrel2}) are discussed in Appendix \ref{obsrelations}. Substituting Eqs.~(\ref{phenrel1}, \ref{phenrel2}) into Eq.~(\ref{fgapprox2}) reduces $\fganal$ to a pure function of $\mg$ and $T$ of the form
\bea
  \log\big[\fganal(\mg,T)\big] & = & \log\big[\fganal(\mg,10)\big] \label{eqfganal2}\\
  && +\delta(0.16\,T-1.6), \nonumber
\eea
where $\fganal(\mg,10)$ is the theoretical \hm/\ha-ratio of a pure disc galaxy, i.e.~$T=10$. The function $\fganal(\mg,10)$ is displayed in Fig.~\ref{fig_etadiscmg} together with the 1-$\sigma$ uncertainty implied by the uncertainties of the four parameters $\alpha_1$, $\alpha_2$, $\gamma_1$, $\gamma_2$. We approximate this relation by the power-law
\be\label{eqfganal3}
  \log\big[\fganal(\mg,10)\big] = c+s\cdot\log\left(\frac{\mg}{10^9\,\h^{-2}\,\msun}\right).
\ee
The parameters minimizing the rms-deviation on the mass-interval $\log(\mg/[\msun\h^{-2}])=7.5-10.5$ are $c=-1.79\pm0.04$ and $s=-0.24\pm0.05$. The given uncertainties approximate the propagated uncertainties of $\alpha_1$, $\alpha_2$, $\gamma_1$, $\gamma_2$.

The simplified theoretical model for the \hm/\ha-ratio given in Eqs.~(\ref{eqfganal2}, \ref{eqfganal3}) exhibits exactly the formal structure of our best phenomenological model 3. Setting $\fganal(\mg,T)$ in Eq.~(\ref{eqfganal2}) equal to $\fgd(\mg,T)$ in Eq.~(\ref{eta3}) for spiral galaxies, yields the following mapping between the theoretical and empirical model-parameters,
\bea\label{match}
  c_3^{\rm sp} & = & c-1.6\,\delta\,, \nonumber \\
  u_3^{\rm sp} & = & s\,, \\
  k_3          & = & 0.16\,\delta\,. \nonumber
\eea

\begin{figure}
  \includegraphics[width=\columnwidth]{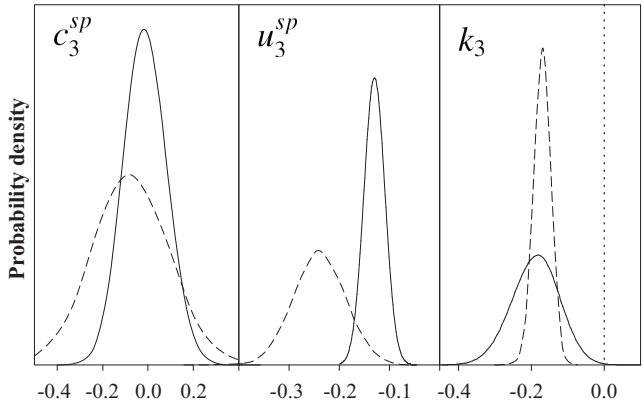}
  \caption{Probability distributions of the three parameters in our model 3 (Eq.~\ref{eta3}) for the \hm/\ha-mass ratio $\fg$ of spiral galaxies. Solid lines represent phenomenologically determined probability distributions given in Table \ref{table_parameters}; dashed lines represent the corresponding theoretical probability distributions, obtained when using Eqs.~(\ref{match}) with the respective distributions for $c$, $s$, and $\delta$.}\label{fig_parametercomparison}
\end{figure}

The probability distributions of the empirical model-parameters on the left hand side of Eqs.~(\ref{match}) were derived in Section \ref{approach2} and their 1-$\sigma$ uncertainties are given in Table \ref{table_parameters}. The corresponding probability distributions of the theoretical model-parameters on the right hand side of Eqs.~(\ref{match}) can be estimated from the Gaussian uncertainties given for the parameters $c$, $s$, $\delta$. The empirical and theoretical parameter distributions are compared in Fig.~\ref{fig_parametercomparison} and reveal a surprising consistency.

\section{Discussion}\label{discussion}

\subsection{Theoretical versus phenomenological model}\label{sec_thandph}

The dependence of $\fg$ on galaxy Hubble type $T$ and cold gas mass $\mg$ was first considered on a purely phenomenological level, and described by the empirical models in Section \ref{approach2}. The best empirical model for spiral galaxies could be quantitatively reproduced by the subsequently derived theoretical model for regular galaxies in Section \ref{local}. Hence, the latter provides a tool for understanding the variations of $\fg$.

In fact, according to Eq.~(\ref{fgapprox2}), $\fg$ seems most directly dictated by the scale radius $\rd$ and the masses $\mg$ and $\msdisk$. The dependence of $\fg$ on $T$ is clearly due to the trend for smaller values of $\rd$ (for a given mass) in bulge-rich galaxies. Several physical reasons for the influence of the bulge on $\rd$ are mentioned in Appendix \ref{scalelengths}.

From Eq.~(\ref{fgapprox2}), one might naively expect that $\fg$ and $\mg$ are positively correlated. However, the disc scale radius $\rd$ increases with $\mg$ as $\rd\propto\mg^{0.66}$ by virtue of Eqs.~(\ref{phenrel1}, \ref{phenrel2}). Taking this scaling into account, the \hm/\ha-ratio $\fg$ effectively decreases with increasing $\mg$. The physical picture is that more massive galaxies are less dense due to their larger sizes, and hence their molecular fraction is lower.

The `best' phenomenological model is by definition the one that, when applied to the galaxies in the HIPASS sample, exhibits the \hm-MF that best fits the reference \hm-MF derived from the CO-LF. The close agreement between the best model defined in this way and the theoretical model therefore supports the accuracy of the CO-LF \citep{Keres2003}, which could a priori be affected by the poorly characterized completeness of the CO-sample. Confirmingly, \cite{Keres2003} argued that the CO-LF does not substantially suffer from incompleteness by analyzing the FIR-LF produced from the same sample.

\subsection{Brief word on cosmic evolution}

The theoretical model $\fganal$ given in Eqs.~(\ref{eqfganal}, \ref{eqfc}) potentially extends to high redshift, as it only premises the invariance of the relation between pressure and $\f$ and a few assumptions with weak dependence on redshift \citep[but see discussion in][]{Obreschkow2009b}. However, we emphasize that the transition from the theoretical model $\fganal$ to the phenomenological model $\fgd$ uses a set of relations extracted from observations in the local Universe. Most probably $\fgd$ underestimates the molecular fraction at higher redshift, predominantly due to the evolution in the mass--diameter relation of Eq.~(\ref{phenrel2}). Indeed, scale radii are smaller at higher redshift for identical masses, thus increasing the pressure and molecular fraction. \citet{Bouwens2004} found $\rd\propto(1+z)^{-1}$ from observations in the Ultra Deep Field, consistent with the theoretical prediction by \citet{Mo1998}. According to Eq.~(\ref{fgapprox2}), where $\fg\propto\rd^{-2.6}$, this implies $\fg\propto(1+z)^{2.6}$. In other words, the phenomenological model 3 (Eq.~\ref{eta3}) for the \hm/\ha-mass ratio should be multiplied by roughly a factor $(1+z)^{2.6}$. However, this conclusion only applies if we consider galaxies with constant stellar and gas masses. For the cosmic evolution of the universal \hm/\ha-ratio $\fu$, we also require a model for the evolution of the stellar and gas mass functions, and it may even be important to consider different scenarios for the evolution of the scale radius for different masses. A more elaborate model for the evolution of $\fu$ can be obtained from cosmological simulations (e.g.~\citealp{Obreschkow2009b} and forthcoming publications).

\subsection{Application: The local cold gas-MF}

We finally apply our best phenomenological model for the \hm/\ha-mass ratio (i.e.~$\fgd$ given in Eq.~\ref{eta3}) to derive an integral cold gas-MF (\ha+\hm+He) from the HIPASS catalog. In fact, the cold gas-MF cannot be inferred solely from the \ha-MF \citep[e.g.][]{Zwaan2005} and the \hm-MF (e.g.~Section \ref{approach1}), but only from a sample of galaxies with simultaneous \ha- and \hm-data. Presently, there is no such sample with a large number of galaxies and an accurate completeness function. Therefore, we prefer using the HIPASS data, which have both sufficient size (4315 galaxies) and well described completeness \citep{Zwaan2004}, and we estimate the corresponding \hm-masses using our model $\fgd$. Details of the computation of the \hm-masses were given in Section \ref{modelevidence}.

\begin{figure}
  \includegraphics[width=\columnwidth]{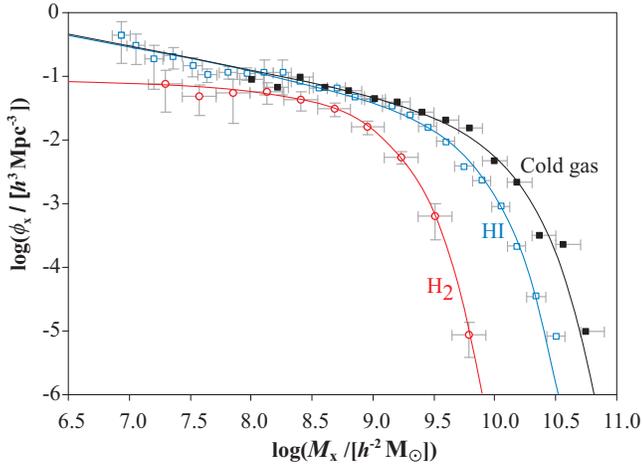}
  \caption{Filled squares represent the integral cold gas-MF (\ha+\hm+He) derived from the HIPASS data using our best phenomenological model for the \hm/\ha-mass ratio (Eq.~\ref{eta3}); empty squares represent the observed \ha-MF \citep{Zwaan2005} and empty circles represent our best estimate of the \hm-MF (Section \ref{approach1}). Solid lines are best fitting Schechter functions.}\label{fig_cold_gas_mf}
\end{figure}

The resulting cold gas-MF is shown in Fig.~\ref{fig_cold_gas_mf} together with the \ha-MF from \cite{Zwaan2005} and the reference \hm-MF derived in Section \ref{approach1}. The displayed continuous functions are best fitting Schechter functions. The respective Schechter function parameters for the cold gas-MF are $\mass^\ast=7.21\cdot10^9\,\h^{-2}\,\msun$, $\alpha=-1.37$, and $\phi^\ast=0.0114\,\h^3\ $Mpc$^{-3}$. The total cold gas density in the local Universe derived by integrating this Schechter function is $\Omegag=4.2\cdot10^{-4}\,\h^{-1}$, closely matching the value $(4.4\pm0.8)\cdot10^{-4}\,\h^{-1}$ obtained when summing up the empirical \ha-density \citep{Zwaan2005}, the \hm-density (Section \ref{approach1}), and the corresponding He-density.

\section{Conclusion}\label{conclusion}

In this paper, we established a coherent picture of the \hm/\ha-ratio in galaxies based on a variety of extragalactic observations and theoretical considerations. Some important jigsaw pieces are:

\begin{enumerate}
  \item Measurements of the $X$-factor \citep[summarized in][]{Boselli2002} were combined with more recent CO-flux measurements and extinction-corrected optical $\magb$-magnitudes, resulting in a working model for $X$.
  \item This model for $X$ was applied to the CO-LF by \citet{Keres2003} in order to derive the first local \hm-MF based on a variable $X$-factor.
  \item Nine samples of local galaxies (245 objects in total) with simultaneous measurements of $\mha$ and $L_{\rm CO}$ were combined to fit a set of empirical models for galactic \hm/\ha-mass ratios $\fg$.
  \item These models were applied to the large \ha-sample of the HIPASS catalog, which permitted the derivation of a \hm-MF for each model for $\fg$. A comparison of these \hm-MFs with the one derived directly from the CO-LF allowed us to determine the statistical evidence of each model and to uncover a clear `best model'.
  \item Based on the relation between pressure and the local \hm/\ha-ratio $\f$ \citep{Leroy2008}, we established a theoretical model for the \hm/\ha-ratio $\fg$ of regular galaxies, which potentially extends to high redshifts.
  \item We could show that the best empirical model for $\fg$ found before is an excellent approximation of the theoretical model in the local Universe.
\end{enumerate}

The factual results standing out of this analysis are

\begin{enumerate}
  \item an empirical \hm-MF obtained by combining the CO-LF of \cite{Keres2003} with a variable $X$-factor (see Fig.~\ref{fig_h2_mf} and parameters in Table \ref{tab_h2mfs}),
  \item an empirical model for $\fg$ (Eq.~\ref{eta3}), which accurately reproduces the above \hm-MF, when applied to the \ha-sample of the HIPASS catalog,
  \item a theoretical model for $\fg$ (Eqs.~\ref{eqfganal}, \ref{eqfc}), which provides a source for physical understanding and generalizes to high redshift,
  \item a quasi-empirical integral cold gas-MF (\ha+\hm+He) based on the HIPASS data.
\end{enumerate}

Self-consistency argues in favour of the interlinked picture established in this paper. However, all quantitative results remain subjected to the uncertainties of the $X$-factor. The latter appears as a scaling factor, affecting in the same way the reference \hm-MF derived from the CO-LF, the phenomenological models of $\fg$ and hence the \hm-MFs derived from HIPASS, as well as the $P$--$\f$ relation and thus the theoretical model for $\fg$. In the future it may therefore be necessary to re-scale the quantitative results of this paper using a more accurate determination of $X$.

\section*{Acknowledgements}
This effort/activity is supported by the European Community Framework Programme 6, Square Kilometre Array Design Studies (SKADS), contract no 011938. We further acknowledge the usage of the HyperLeda database (http://leda.univ-lyon1.fr) and we thank the anonymous referee for the helpful suggestions.

%\bibliography{../Bibliography/astro}
%\bibliographystyle{../Bibliography/my_mn2e}

\appendix

\section{Homogenized data}\label{data}
This section presents the data (245 galaxies) used for the derivation of the models of $\fg$ in section~\ref{approach2}.

CO-luminosities were drawn from 10 smaller samples: 17 nearby ($\lesssim10\ $Mpc) lenticulars and ellipticals \citep{Welch2003,Sage2006}, 4 late-type spirals \citep{Matthews2005}, 68 isolated late-type spirals \citep{Sauty2003}, 6 ellipticals \citep{Georgakakis2001}, 17 spirals of all types \citep{Andreani1995}, 48 nearby ($\lesssim10\ $Mpc) spirals of all types \citep{Sage1993}, 12 ellipticals \citep{Lees1991}, 18 lenticulars and ellipticals \citep{Thronson1989}, 77 spirals of all types \citep{Young1989b}. These 267 objects contained 22 repeated galaxies. In each case of repetition, the older reference was removed, such as to remain with the 245 distinct sources listed in Table \ref{table_data}. The CO-luminosities were homogenized by making them independent of different $X$-factors and Hubble constants. All other properties listed in the table were taken from homogenized reference catalogs, such as described in Section \ref{observedsample}.

\onecolumn

\begin{longtable}{lccccccc}

\caption{Homogenized galaxy sample based on data drawn from the literature. $T$ is the numerical Hubble type (see online help of the HyperLeda database), $D_{\rm l}$ the luminosity distance, $\magb$ is the extinction corrected absolute blue magnitude, and $X$ is the variable $X$-factor derived from $\magb$ (Eq.~\ref{xmb}) without addition of Gaussian scatter. The references for \hm-masses are: [1] \citet{Welch2003,Sage2006}, [2] \citet{Matthews2005}, [3] \citet{Sauty2003}, [4] \citet{Georgakakis2001}, [5] \citet{Andreani1995}, [6] \citet{Sage1993}, [7] \citet{Lees1991}, [8] \citet{Thronson1989}, [9] \citet{Young1989b}.}

\label{table_data} \\

%This is the header for the first page of the table...
   \hline \\ [-2ex]
   \multicolumn{1}{c}{Object} &
   \multicolumn{1}{c}{$T$} &
   \multicolumn{1}{c}{$D_{\rm l}\,/\,\h^{-1}\,$Mpc} &
   \multicolumn{1}{c}{$\magb-5\,\log\,\h$} & %
   \multicolumn{1}{c}{$X$} &
   \multicolumn{1}{c}{$\log(\mhm/X\,\h^{-2}\,\msun)$} &
   \multicolumn{1}{c}{Ref.~\hm} &
   \multicolumn{1}{c}{$\log(\mha/\h^{-2}\,\msun)$} \\
   \hline \\ [-2ex]
\endfirsthead

%This is the header for the remaining page(s) of the table...
   \hline \\ [-2ex]
   \multicolumn{1}{c}{Object} &
   \multicolumn{1}{c}{$T$} &
   \multicolumn{1}{c}{$D_{\rm l}\,/[\,\h^{-1}\,$Mpc]} &
   \multicolumn{1}{c}{$\magb-5\,\log\,\h$} & %
   \multicolumn{1}{c}{$X$} &
   \multicolumn{1}{c}{$\log(\mhm/[X\,\h^{-2}\,\msun])$} &
   \multicolumn{1}{c}{Ref.~\hm} &
   \multicolumn{1}{c}{$\log(\mha/[\h^{-2}\,\msun])$} \\
   \hline \\ [-2ex]
\endhead

% Data rows
   NGC 404 &       -2.8 &        1.7 &     -15.86 &       6.66 &       6.06 &          1 &       7.51 \\
  NGC 2787 &       -1.1 &        9.5 &     -18.87 &       2.21 &       6.58 &          1 &       8.58 \\
  NGC 3115 &       -2.8 &        6.4 &     -19.27 &       1.91 &       5.60 &          1 &       6.75 \\
  NGC 3384 &       -2.7 &        9.2 &     -19.06 &       2.06 &       5.87 &          1 &       5.94 \\
  NGC 3489 &       -1.3 &        7.7 &     -18.45 &       2.58 &       6.12 &          1 &       6.46 \\
  NGC 3607 &       -3.1 &       10.3 &     -19.23 &       1.94 &       8.34 &          1 &       6.93 \\
  NGC 3870 &       -2.0 &        9.9 &     -16.56 &       5.15 &       7.44 &          1 &       8.08 \\
  NGC 3941 &       -2.0 &       11.0 &     -19.04 &       2.08 &       7.15 &          1 &       8.81 \\
  NGC 4026 &       -1.8 &       12.1 &     -18.82 &       2.25 &       7.27 &          1 &       7.86 \\
  NGC 4150 &       -2.1 &        6.8 &     -17.66 &       3.44 &       6.91 &          1 &       6.88 \\
  NGC 4203 &       -2.7 &       12.7 &     -18.86 &       2.22 &       6.21 &          1 &       8.41 \\
  NGC 4310 &       -1.0 &       10.8 &     -16.86 &       4.61 &       6.96 &          1 &       7.10 \\
  NGC 4460 &       -0.9 &        7.3 &     -17.04 &       4.32 &       6.45 &          1 &       8.26 \\
  NGC 4880 &       -1.5 &       14.8 &     -17.92 &       3.13 &       6.27 &          1 &       6.02 \\
  NGC 7013 &        0.5 &        9.6 &     -18.79 &       2.28 &       7.30 &          1 &       8.70 \\
  NGC 7077 &       -3.9 &       12.0 &     -16.13 &       6.03 &       6.09 &          1 &       7.60 \\
  NGC 7457 &       -2.6 &        9.6 &     -18.29 &       2.73 &       5.85 &          1 &       5.88 \\
   NGC 100 &        5.9 &        9.0 &     -17.61 &       3.51 &       5.91 &          2 &       8.87 \\
  UGC 2082 &        5.9 &        7.7 &     -17.72 &       3.37 &       5.89 &          2 &       8.80 \\
  UGC 3137 &        4.2 &       12.5 &     -17.05 &       4.30 &       6.20 &          2 &       9.11 \\
  UGC 6667 &        6.0 &       12.1 &     -17.06 &       4.29 &       5.73 &          2 &       8.54 \\
     UGC 5 &        3.9 &       74.4 &     -20.98 &       1.02 &       8.76 &          3 &       9.82 \\
  NGC 7817 &        4.1 &       24.1 &     -20.42 &       1.25 &       8.44 &          3 &       9.30 \\
   IC 1551 &        3.6 &      136.0 &     -22.17 &       0.66 &       8.94 &          3 &       9.34 \\
   NGC 237 &        4.5 &       42.0 &     -19.81 &       1.57 &       8.53 &          3 &       9.75 \\
   NGC 575 &        5.3 &       32.3 &     -19.10 &       2.03 &       8.04 &          3 &       9.18 \\
   NGC 622 &        3.4 &       52.1 &     -19.93 &       1.50 &       8.24 &          3 &       9.54 \\
  UGC 1167 &        5.9 &       43.6 &     -19.18 &       1.97 &       8.85 &          3 &       9.61 \\
  UGC 1395 &        3.1 &       52.3 &     -19.90 &       1.51 &       8.43 &          3 &       9.25 \\
  UGC 1587 &        3.7 &       57.4 &     -20.38 &       1.27 &       7.86 &          3 &       9.59 \\
  UGC 1706 &        5.8 &       49.4 &     -19.82 &       1.56 &       7.96 &          3 &       9.17 \\
    IC 302 &        4.1 &       59.6 &     -21.33 &       0.90 &       8.43 &          3 &      10.19 \\
    IC 391 &        4.9 &       18.3 &     -18.91 &       2.18 &       7.46 &          3 &       8.89 \\
  UGC 3420 &        3.1 &       54.5 &     -20.96 &       1.03 &       8.03 &          3 &      10.01 \\
  UGC 3581 &        5.2 &       53.2 &     -20.30 &       1.31 &       8.24 &          3 &       9.56 \\
  NGC 2344 &        4.4 &       11.3 &     -17.91 &       3.14 &       6.73 &          3 &       8.66 \\
  UGC 3863 &        1.1 &       62.2 &     -20.53 &       1.20 &       8.32 &          3 &       9.30 \\
  UGC 4684 &        7.2 &       24.9 &     -17.92 &       3.13 &       6.82 &          3 &       9.11 \\
  NGC 2746 &        1.1 &       73.7 &     -20.65 &       1.15 &       8.65 &          3 &       9.64 \\
  UGC 4781 &        5.9 &       14.4 &     -16.54 &       5.19 &       6.46 &          3 &       8.90 \\
  UGC 5055 &        3.1 &       79.4 &     -20.19 &       1.36 &       8.79 &          3 &      10.02 \\
  NGC 2900 &        5.9 &       54.3 &     -19.51 &       1.75 &       8.57 &          3 &       9.69 \\
  NGC 2977 &        3.2 &       33.5 &     -19.95 &       1.49 &       8.31 &          3 &       8.83 \\
  NGC 3049 &        2.5 &       15.0 &     -17.86 &       3.20 &       7.24 &          3 &       8.86 \\
    IC 651 &        8.2 &       45.2 &     -20.37 &       1.28 &       8.54 &          3 &       9.53 \\
  NGC 3526 &        5.2 &       14.5 &     -18.68 &       2.37 &       7.73 &          3 &       8.64 \\
  UGC 6568 &        8.2 &       60.8 &     -19.86 &       1.54 &       8.12 &          3 &       9.14 \\
  UGC 6769 &        3.0 &       88.2 &     -20.66 &       1.15 &       9.10 &          3 &       9.96 \\
  UGC 6780 &        6.4 &       17.3 &     -16.79 &       4.73 &       7.29 &          3 &       9.28 \\
  UGC 6879 &        7.1 &       24.1 &     -18.78 &       2.28 &       7.82 &          3 &       8.83 \\
  UGC 6903 &        5.9 &       19.3 &     -17.69 &       3.40 &       7.46 &          3 &       9.07 \\
  NGC 4348 &        4.1 &       20.3 &     -19.49 &       1.76 &       8.10 &          3 &       9.01 \\
  NGC 4617 &        3.1 &       49.6 &     -20.70 &       1.13 &       8.56 &          3 &       9.90 \\
  NGC 4635 &        6.5 &       10.9 &     -17.28 &       3.96 &       6.73 &          3 &       8.23 \\
  NGC 5377 &        1.1 &       20.6 &     -19.83 &       1.55 &       7.81 &          3 &       8.91 \\
  NGC 5375 &        2.4 &       26.0 &     -19.54 &       1.73 &       7.60 &          3 &       9.24 \\
  NGC 5584 &        5.9 &       17.1 &     -19.06 &       2.06 &       7.22 &          3 &       9.27 \\
  NGC 5690 &        5.4 &       18.4 &     -19.88 &       1.53 &       8.15 &          3 &       9.33 \\
  NGC 5768 &        5.3 &       20.3 &     -18.74 &       2.32 &       7.90 &          3 &       9.11 \\
  NGC 5772 &        3.1 &       52.3 &     -20.41 &       1.26 &       8.25 &          3 &       9.49 \\
  NGC 5913 &        1.3 &       20.8 &     -19.00 &       2.11 &       8.22 &          3 &       8.44 \\
  NGC 6012 &        1.9 &       20.1 &     -19.00 &       2.11 &       7.73 &          3 &       9.26 \\
   IC 1231 &        5.8 &       55.9 &     -20.71 &       1.13 &       8.04 &          3 &       9.14 \\
 UGC 10699 &        4.4 &       65.5 &     -20.19 &       1.36 &       8.60 &          3 &       9.11 \\
 UGC 10743 &        1.1 &       27.2 &     -18.75 &       2.31 &       7.52 &          3 &       8.78 \\
  NGC 6347 &        3.1 &       64.3 &     -20.46 &       1.23 &       8.57 &          3 &       9.48 \\
 UGC 10862 &        5.3 &       18.2 &     -17.81 &       3.26 &       7.21 &          3 &       9.07 \\
  NGC 6389 &        3.6 &       33.1 &     -20.37 &       1.28 &       8.30 &          3 &       9.93 \\
 UGC 11058 &        3.2 &       50.6 &     -20.48 &       1.22 &       8.51 &          3 &       9.40 \\
  NGC 6643 &        5.2 &       17.8 &     -20.31 &       1.30 &       8.35 &          3 &       9.27 \\
  NGC 6711 &        4.0 &       50.1 &     -20.18 &       1.37 &       8.77 &          3 &       9.14 \\
 UGC 11635 &        3.7 &       51.8 &     -21.05 &       0.99 &       8.95 &          3 &       9.88 \\
 UGC 11723 &        3.1 &       50.1 &     -19.87 &       1.53 &       8.48 &          3 &       9.57 \\
  NGC 7056 &        3.6 &       55.8 &     -20.53 &       1.20 &       8.67 &          3 &       9.11 \\
  NGC 7156 &        5.9 &       40.8 &     -20.12 &       1.40 &       8.43 &          3 &       9.32 \\
 UGC 11871 &        3.1 &       82.9 &     -20.38 &       1.27 &       9.22 &          3 &       9.43 \\
  NGC 7328 &        2.1 &       29.2 &     -19.31 &       1.88 &       8.34 &          3 &       9.45 \\
  NGC 7428 &        1.1 &       31.0 &     -18.85 &       2.23 &       7.72 &          3 &       9.44 \\
 UGC 12304 &        5.2 &       35.3 &     -19.40 &       1.82 &       8.01 &          3 &       8.88 \\
 UGC 12372 &        4.0 &       57.7 &     -19.94 &       1.49 &       8.65 &          3 &       9.49 \\
  NGC 7514 &        4.3 &       51.1 &     -20.62 &       1.16 &       8.18 &          3 &       9.16 \\
 UGC 12474 &        1.1 &       53.5 &     -20.53 &       1.20 &       8.80 &          3 &       8.87 \\
  NGC 7664 &        5.1 &       36.3 &     -20.03 &       1.44 &       8.51 &          3 &       9.91 \\
 UGC 12646 &        3.0 &       83.7 &     -20.84 &       1.07 &       8.68 &          3 &       9.70 \\
  NGC 7712 &        1.6 &       31.9 &     -18.94 &       2.15 &       7.84 &          3 &       9.10 \\
   IC 1508 &        7.2 &       43.8 &     -20.07 &       1.42 &       8.45 &          3 &       9.75 \\
 UGC 12776 &        3.0 &       51.8 &     -19.88 &       1.53 &       8.31 &          3 &       9.99 \\
   IC 5355 &        5.7 &       50.8 &     -19.56 &       1.72 &       8.26 &          3 &       9.05 \\
 UGC 12840 &       -1.8 &       71.3 &     -20.27 &       1.32 &       7.97 &          3 &       9.43 \\
  NGC 2623 &        2.0 &       57.2 &     -20.59 &       1.18 &       9.02 &          4 &       9.01 \\
  NGC 2865 &       -4.1 &       26.0 &     -20.01 &       1.46 &       7.35 &          4 &       8.79 \\
  NGC 3921 &        0.0 &       61.9 &     -21.00 &       1.01 &       8.82 &          4 &       9.46 \\
  NGC 4649 &       -4.6 &       12.1 &     -20.70 &       1.13 &       7.15 &          4 &       8.35 \\
  NGC 7252 &       -2.1 &       47.0 &     -20.73 &       1.12 &       8.83 &          4 &       9.29 \\
  NGC 7727 &        1.1 &       17.9 &     -19.98 &       1.47 &       7.27 &          4 &       8.45 \\
   NGC 142 &        3.1 &       81.4 &     -20.46 &       1.23 &       9.36 &          5 &       9.43 \\
   IC 1553 &        5.4 &       28.0 &     -18.75 &       2.31 &       7.69 &          5 &       9.10 \\
ESO 473-27 &        4.4 &      193.6 &     -21.12 &       0.97 &       9.78 &          5 &       9.75 \\
   NGC 232 &        1.1 &       66.7 &     -19.82 &       1.56 &       9.50 &          5 &       9.21 \\
ESO 475-16 &        2.1 &       70.7 &     -20.44 &       1.24 &       9.01 &          5 &       9.74 \\
   NGC 578 &        5.0 &       14.8 &     -19.73 &       1.61 &       7.97 &          5 &       9.52 \\
 ESO 478-6 &        4.1 &       52.6 &     -20.64 &       1.16 &       8.96 &          5 &       9.23 \\
  NGC 1187 &        5.0 &       12.2 &     -19.39 &       1.83 &       8.69 &          5 &       9.33 \\
  NGC 1306 &        2.8 &       12.7 &     -16.85 &       4.63 &       7.34 &          5 &       8.64 \\
  NGC 1385 &        5.9 &       13.1 &     -19.56 &       1.72 &       8.59 &          5 &       9.07 \\
ESO 549-23 &        1.2 &       40.8 &     -19.42 &       1.81 &       8.45 &          5 &       8.88 \\
ESO 483-12 &        0.3 &       41.0 &     -19.18 &       1.97 &       8.27 &          5 &       8.83 \\
  NGC 1591 &        1.9 &       39.5 &     -19.65 &       1.66 &       8.51 &          5 &       9.04 \\
  NGC 7115 &        3.4 &       34.1 &     -19.53 &       1.73 &       8.26 &          5 &       9.52 \\
  NGC 7225 &       -0.5 &       47.9 &     -20.09 &       1.41 &       9.29 &          5 &       9.07 \\
  NGC 7314 &        4.0 &       13.2 &     -19.71 &       1.62 &       8.05 &          5 &       9.24 \\
   NGC 628 &        5.2 &        6.9 &     -19.84 &       1.55 &       8.55 &          6 &       9.73 \\
   NGC 672 &        6.0 &        5.1 &     -19.03 &       2.08 &       6.60 &          6 &       9.07 \\
   NGC 891 &        3.0 &        6.7 &     -19.43 &       1.80 &       8.97 &          6 &       9.72 \\
   NGC 925 &        7.0 &        6.6 &     -19.32 &       1.87 &       8.04 &          6 &       9.57 \\
  NGC 1058 &        5.3 &        6.3 &     -17.78 &       3.29 &       7.42 &          6 &       8.93 \\
  NGC 1560 &        7.0 &        2.3 &     -15.91 &       6.53 &       5.88 &          6 &       8.47 \\
  NGC 2403 &        6.0 &        3.2 &     -18.89 &       2.19 &       7.31 &          6 &       9.54 \\
  NGC 2683 &        3.1 &        5.2 &     -19.53 &       1.73 &       7.63 &          6 &       8.54 \\
  NGC 2903 &        4.0 &        6.3 &     -20.16 &       1.38 &       8.39 &          6 &       9.01 \\
  NGC 2976 &        5.3 &        1.6 &     -17.35 &       3.86 &       6.42 &          6 &       7.49 \\
  NGC 3031 &        2.4 &        2.4 &     -19.90 &       1.51 &       7.42 &          6 &       9.15 \\
  NGC 3184 &        5.9 &        7.7 &     -19.11 &       2.02 &       8.35 &          6 &       9.11 \\
  NGC 3344 &        4.0 &        6.9 &     -18.89 &       2.19 &       7.74 &          6 &       9.01 \\
  NGC 3351 &        3.0 &        8.3 &     -19.46 &       1.78 &       8.08 &          6 &       8.67 \\
  NGC 3368 &        1.8 &        9.4 &     -20.12 &       1.40 &       8.18 &          6 &       8.95 \\
  NGC 3486 &        5.2 &        8.2 &     -18.84 &       2.23 &       7.50 &          6 &       9.03 \\
  NGC 3521 &        4.0 &        8.0 &     -20.31 &       1.30 &       8.75 &          6 &       9.63 \\
  NGC 3593 &       -0.4 &        6.9 &     -17.50 &       3.65 &       7.62 &          6 &       7.75 \\
  NGC 3623 &        1.0 &        8.9 &     -20.17 &       1.37 &       7.62 &          6 &       8.27 \\
  NGC 3627 &        3.0 &        7.9 &     -20.40 &       1.26 &       8.55 &          6 &       8.56 \\
  NGC 3628 &        3.1 &        9.2 &     -20.67 &       1.14 &       8.62 &          6 &       9.33 \\
  NGC 4020 &        6.9 &        9.2 &     -17.31 &       3.91 &       6.60 &          6 &       8.05 \\
  NGC 4062 &        5.3 &        9.4 &     -18.78 &       2.28 &       7.63 &          6 &       8.47 \\
  NGC 4096 &        5.3 &        7.9 &     -19.49 &       1.76 &       7.75 &          6 &       8.86 \\
  NGC 4144 &        6.0 &        3.1 &     -15.93 &       6.48 &       6.31 &          6 &       8.09 \\
  NGC 4244 &        6.1 &        2.3 &     -18.06 &       2.97 &       6.62 &          6 &       8.72 \\
  NGC 4245 &        0.1 &       10.5 &     -17.97 &       3.07 &       7.39 &          6 &       6.61 \\
  NGC 4274 &        1.7 &       10.9 &     -19.33 &       1.87 &       8.27 &          6 &       8.75 \\
  NGC 4288 &        7.1 &        7.5 &     -16.32 &       5.62 &       6.67 &          6 &       8.52 \\
  NGC 4314 &        1.0 &       11.5 &     -19.02 &       2.09 &       7.69 &          6 &       6.43 \\
  NGC 4359 &        5.0 &       14.3 &     -17.49 &       3.66 &       6.55 &          6 &       8.44 \\
  NGC 4414 &        5.1 &        8.9 &     -19.25 &       1.92 &       8.48 &          6 &       8.90 \\
  NGC 4448 &        1.8 &        8.2 &     -17.86 &       3.20 &       7.39 &          6 &       7.38 \\
  NGC 4490 &        7.0 &        8.0 &     -20.93 &       1.04 &       7.45 &          6 &       9.54 \\
  NGC 4437 &        6.0 &       11.6 &     -20.70 &       1.13 &       8.14 &          6 &       7.90 \\
  NGC 4525 &        5.9 &       13.5 &     -18.11 &       2.92 &       6.57 &          6 &       7.86 \\
  NGC 4559 &        6.0 &        9.8 &     -20.35 &       1.28 &       8.26 &          6 &       9.57 \\
  NGC 4565 &        3.2 &       13.8 &     -21.74 &       0.77 &       8.62 &          6 &       9.48 \\
  NGC 4605 &        4.9 &        3.0 &     -17.58 &       3.54 &       6.82 &          6 &       8.05 \\
  NGC 4631 &        6.6 &        7.9 &     -21.46 &       0.86 &       8.03 &          6 &       9.58 \\
  NGC 4736 &        2.4 &        5.3 &     -19.27 &       1.91 &       7.86 &          6 &       8.23 \\
  NGC 4826 &        2.4 &        5.5 &     -19.86 &       1.54 &       7.79 &          6 &       8.07 \\
  NGC 5055 &        4.0 &        7.3 &     -20.43 &       1.25 &       8.80 &          6 &       9.40 \\
  NGC 5194 &        4.0 &        7.0 &     -19.74 &       1.61 &       9.29 &          6 &       9.21 \\
  NGC 5457 &        5.9 &        5.0 &     -20.26 &       1.33 &       8.50 &          6 &       9.79 \\
  NGC 6503 &        5.9 &        4.6 &     -17.77 &       3.31 &       7.35 &          6 &       8.86 \\
  NGC 6946 &        5.9 &        4.1 &     -20.12 &       1.40 &       8.74 &          6 &       9.55 \\
  NGC 7640 &        5.3 &        5.5 &     -18.75 &       2.31 &       6.93 &          6 &       9.62 \\
   NGC 185 &       -4.8 &        0.7 &     -13.83 &      14.00 &       4.81 &          7 &       5.18 \\
   NGC 205 &       -4.7 &        0.7 &     -13.61 &      15.18 &       4.95 &          7 &       5.57 \\
   NGC 855 &       -4.6 &        6.9 &     -16.23 &       5.81 &       5.33 &          7 &       7.62 \\
  NGC 3265 &       -4.8 &       15.7 &     -17.28 &       3.96 &       7.13 &          7 &       7.95 \\
  NGC 3928 &       -4.5 &       12.1 &     -17.35 &       3.86 &       7.36 &          7 &       8.22 \\
  NGC 5128 &       -2.1 &        5.3 &     -20.59 &       1.17 &       8.16 &          7 &       8.28 \\
  NGC 5666 &        6.4 &       23.6 &     -18.90 &       2.19 &       8.00 &          7 &       8.63 \\
  NGC 1819 &       -1.9 &       44.8 &     -20.23 &       1.34 &       9.10 &          7 &       9.13 \\
  NGC 3032 &       -1.8 &       16.7 &     -18.14 &       2.89 &       7.72 &          7 &       7.76 \\
  NGC 4138 &       -0.9 &       10.9 &     -17.97 &       3.07 &       7.13 &          7 &       8.54 \\
  NGC 7465 &       -1.9 &       20.6 &     -18.57 &       2.47 &       8.11 &          7 &       9.20 \\
  NGC 3413 &       -1.8 &        7.9 &     -16.66 &       4.96 &       7.21 &          8 &       7.95 \\
  NGC 5866 &       -1.2 &        9.5 &     -19.23 &       1.94 &       7.81 &          8 &       8.15 \\
  NGC 4710 &       -0.8 &       13.8 &     -19.02 &       2.09 &       8.25 &          8 &       7.20 \\
  NGC 4459 &       -1.4 &       13.3 &     -19.37 &       1.84 &       8.30 &          8 &       6.70 \\
  NGC 4526 &       -1.9 &        6.7 &     -18.63 &       2.41 &       8.30 &          8 &       7.05 \\
   NGC 693 &        0.1 &       15.5 &     -18.08 &       2.95 &       7.53 &          8 &       8.74 \\
  NGC 2685 &       -1.1 &       11.0 &     -18.32 &       2.70 &       7.45 &          8 &       8.79 \\
  NGC 2273 &        1.0 &       20.7 &     -19.47 &       1.77 &       8.26 &          8 &       8.90 \\
  NGC 3611 &        1.1 &       16.1 &     -18.67 &       2.38 &       8.42 &          8 &       8.75 \\
  NGC 4457 &        0.4 &        9.4 &     -18.31 &       2.71 &       8.63 &          8 &       8.27 \\
  NGC 4383 &        1.0 &       18.3 &     -19.01 &       2.10 &       7.91 &          8 &       9.15 \\
  NGC 7625 &        1.2 &       17.2 &     -18.38 &       2.64 &       8.56 &          8 &       8.98 \\
    NGC 23 &        1.2 &       47.4 &     -20.84 &       1.07 &       9.30 &          9 &       9.69 \\
   NGC 253 &        5.1 &        1.7 &     -20.19 &       1.36 &       8.32 &          9 &       9.04 \\
   NGC 520 &        0.8 &       21.5 &     -19.90 &       1.51 &       9.35 &          9 &       9.50 \\
   NGC 828 &        1.0 &       55.9 &     -20.95 &       1.03 &       9.75 &          9 &       9.80 \\
   NGC 834 &        3.9 &       48.1 &     -20.32 &       1.30 &       9.13 &          9 &       9.47 \\
   NGC 864 &        5.1 &       15.4 &     -19.80 &       1.57 &       8.49 &          9 &       9.78 \\
   NGC 877 &        4.8 &       39.7 &     -21.15 &       0.96 &       9.34 &          9 &      10.08 \\
  NGC 1055 &        3.2 &        9.3 &     -18.97 &       2.13 &       9.37 &          9 &       9.39 \\
    IC 342 &        5.9 &        2.3 &     -19.85 &       1.54 &       8.70 &          9 &       9.68 \\
  NGC 1530 &        3.1 &       27.5 &     -20.70 &       1.13 &       9.10 &          9 &       9.76 \\
  NGC 1569 &        9.6 &        2.4 &     -15.94 &       6.46 &       5.89 &          9 &       8.09 \\
  NGC 1614 &        4.9 &       47.2 &     -20.64 &       1.16 &       9.36 &          9 &       9.28 \\
  NGC 2146 &        2.3 &       11.6 &     -20.34 &       1.29 &       9.04 &          9 &       9.50 \\
  NGC 2339 &        4.0 &       22.9 &     -20.02 &       1.45 &       9.27 &          9 &       9.45 \\
  NGC 2276 &        5.4 &       27.2 &     -20.80 &       1.09 &       9.31 &          9 &       9.50 \\
  NGC 2532 &        5.2 &       54.5 &     -21.00 &       1.01 &       9.10 &          9 &       9.92 \\
  NGC 2633 &        3.0 &       24.5 &     -19.47 &       1.77 &       8.83 &          9 &       9.41 \\
  NGC 2775 &        1.7 &       13.5 &     -19.83 &       1.55 &       8.30 &          9 &       8.16 \\
  NGC 2841 &        3.0 &        8.3 &     -20.07 &       1.42 &       8.61 &          9 &       9.20 \\
  NGC 3034 &        8.0 &        1.7 &     -17.30 &       3.93 &       8.16 &          9 &       8.54 \\
  NGC 3079 &        6.6 &       13.5 &     -20.68 &       1.14 &       9.16 &          9 &       9.57 \\
  NGC 3147 &        3.9 &       31.1 &     -21.43 &       0.86 &       9.65 &          9 &       9.52 \\
  NGC 3221 &        5.6 &       42.5 &     -19.86 &       1.54 &       9.24 &          9 &       9.81 \\
  NGC 3310 &        4.0 &       12.2 &     -19.26 &       1.92 &       7.81 &          9 &       9.33 \\
  NGC 3437 &        5.2 &       13.8 &     -19.03 &       2.08 &       7.91 &          9 &       9.03 \\
  NGC 3504 &        2.1 &       16.8 &     -19.68 &       1.64 &       8.50 &          9 &       8.37 \\
  NGC 3556 &        6.0 &        9.3 &     -19.89 &       1.52 &       8.37 &          9 &       9.35 \\
  NGC 3893 &        5.1 &       12.0 &     -20.13 &       1.39 &       8.35 &          9 &       9.29 \\
  NGC 4192 &        2.5 &       10.0 &     -20.83 &       1.08 &       8.57 &          9 &       9.33 \\
  NGC 4194 &        9.7 &       27.3 &     -19.87 &       1.53 &       8.61 &          9 &       8.96 \\
  NGC 4254 &        5.2 &       25.2 &     -21.82 &       0.75 &       9.07 &          9 &       9.39 \\
  NGC 4303 &        4.0 &       16.2 &     -21.05 &       0.99 &       8.95 &          9 &       9.38 \\
  NGC 4321 &        4.0 &       16.8 &     -21.29 &       0.91 &       9.12 &          9 &       9.06 \\
  NGC 4388 &        2.8 &       26.2 &     -21.16 &       0.95 &       7.96 &          9 &       8.33 \\
  NGC 4394 &        2.9 &       10.3 &     -18.65 &       2.39 &       8.04 &          9 &       8.22 \\
  NGC 4402 &        3.3 &       10.0 &     -17.18 &       4.09 &       8.39 &          9 &       8.23 \\
  NGC 4419 &        1.1 &       10.0 &     -18.43 &       2.60 &       8.56 &          9 &       7.62 \\
  NGC 4424 &        1.2 &        5.1 &     -15.75 &       6.91 &       7.34 &          9 &       7.89 \\
  NGC 4438 &        0.7 &       10.0 &     -19.99 &       1.47 &       7.92 &          9 &       8.26 \\
  NGC 4449 &        9.8 &        2.6 &     -16.83 &       4.66 &       6.60 &          9 &       9.10 \\
  NGC 4450 &        2.3 &       20.7 &     -21.10 &       0.98 &       8.25 &          9 &       7.95 \\
  NGC 4501 &        3.4 &       23.9 &     -22.33 &       0.62 &       8.94 &          9 &       8.91 \\
  NGC 4527 &        4.0 &       17.9 &     -20.75 &       1.11 &       8.85 &          9 &       9.35 \\
  NGC 4535 &        5.0 &       20.3 &     -21.18 &       0.95 &       8.79 &          9 &       9.41 \\
  NGC 4536 &        4.2 &       18.6 &     -21.02 &       1.01 &       8.46 &          9 &       9.23 \\
  NGC 4548 &        3.1 &        6.0 &     -20.01 &       1.46 &       8.33 &          9 &       8.68 \\
  NGC 4569 &        2.4 &       10.0 &     -20.33 &       1.29 &       8.77 &          9 &       8.32 \\
  NGC 4571 &        6.3 &       10.0 &     -17.53 &       3.60 &       8.17 &          9 &       8.49 \\
  NGC 4579 &        2.8 &       16.1 &     -20.91 &       1.05 &       8.55 &          9 &       8.38 \\
  NGC 4647 &        5.2 &       14.9 &     -19.02 &       2.09 &       8.37 &          9 &       8.33 \\
  NGC 4651 &        5.2 &        9.1 &     -18.86 &       2.22 &       8.14 &          9 &       9.21 \\
  NGC 4654 &        5.9 &       11.4 &     -19.86 &       1.54 &       8.46 &          9 &       9.15 \\
  NGC 4689 &        4.7 &       17.2 &     -19.92 &       1.50 &       8.44 &          9 &       8.30 \\
  NGC 5236 &        5.0 &        4.5 &     -19.99 &       1.46 &       9.41 &          9 &       9.86 \\
  NGC 5936 &        3.2 &       42.2 &     -20.29 &       1.31 &       9.15 &          9 &       8.90 \\
  NGC 6207 &        4.9 &       10.9 &     -19.11 &       2.02 &       7.52 &          9 &       8.97 \\
  NGC 6574 &        3.9 &       24.6 &     -20.12 &       1.40 &       9.13 &          9 &       8.87 \\
  NGC 7217 &        2.5 &       11.2 &     -19.70 &       1.63 &       8.41 &          9 &       8.65 \\
  NGC 7331 &        3.9 &        9.9 &     -20.81 &       1.09 &       9.22 &          9 &       9.77 \\
  NGC 7469 &        1.1 &       50.4 &     -21.00 &       1.01 &       9.50 &          9 &       9.30 \\
  NGC 7479 &        4.4 &       24.6 &     -20.93 &       1.04 &       9.35 &          9 &       9.77 \\
  NGC 7541 &        4.7 &       27.2 &     -20.78 &       1.10 &       9.34 &          9 &      10.01 \\
  NGC 7674 &        3.8 &       91.6 &     -21.17 &       0.95 &       9.66 &          9 &      10.11 \\
\end{longtable}

\twocolumn

\section{Diverse phenomenological relations}\label{obsrelations}

This section summarizes the two phenomenological relations given in Eqs.~(\ref{phenrel1}) and (\ref{phenrel2}).

\subsection{Stellar mass versus gas mass}\label{stvsgs}

From the galaxy sample presented in Appendix \ref{data}, we extracted all 25 Scd/Sd-type galaxies ($6\leq T\leq9$), that is all objects approximating pure discs. For these objects the total gas masses $\mg$ were calculated via $\mg=(\mha+\mhm)/\hfraction$. Additionally, we estimated the stellar mass $\ms$ of each galaxy from the I-band magnitude $M_{\rm I}$ via \citep{Mo1998},
\be\label{eqms}
  \log(\ms/\msun) = 1.66+\log(\Upsilon_{\rm I})-M_{\rm I}/2.5,
\ee
where the mass/light-ratio $\log(\Upsilon_{\rm I})=1.2$ has been adopted from \cite{McGaugh1997}.

\begin{figure}
  \includegraphics[width=\columnwidth]{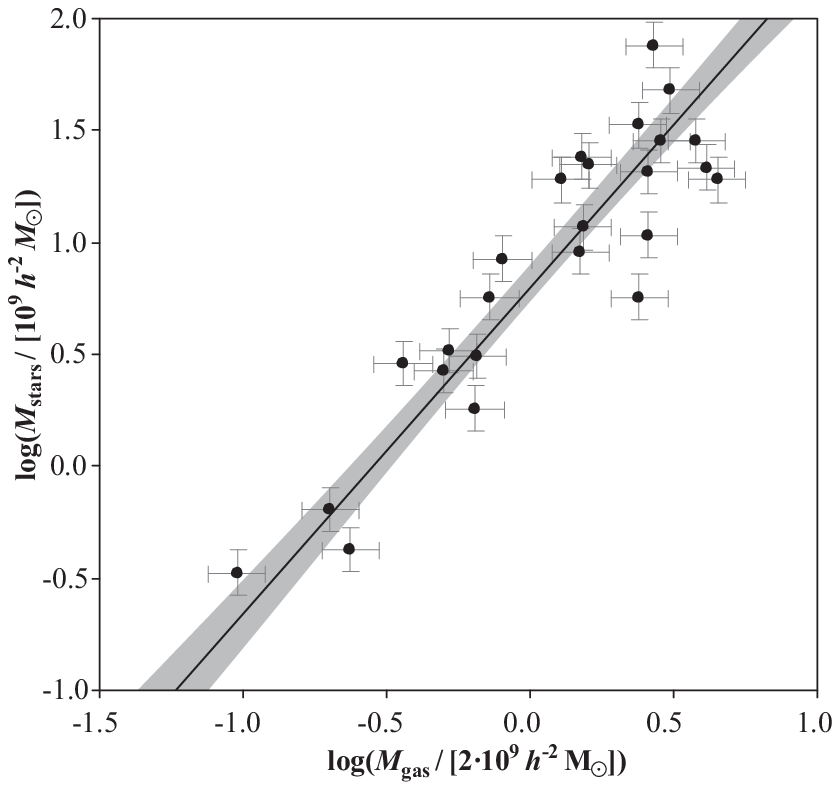}
  \caption{Data points (subsample of the data shown in Appendix \ref{data}) represent the observed relation between disc stellar mass $\msdisk$ and cold gas mass $\mg$. The solid line shows the best power-law fit and the shaded envelope its 1-$\sigma$ uncertainty. This power-law is given in Eq.~(\ref{phenrel1}) and has a slope of $\alpha_1=1.46\pm0.10$.}
  \label{fig_phenrel1}
\end{figure}

The resulting data points displayed in Fig.~\ref{fig_phenrel1} reveal an approximate power-law relation between $\mg$ and $\ms$. We have fitted the corresponding free parameters $\alpha_1,\gamma_1$ to the data points by minimizing the x-y-weighted rms-deviations. 1-$\sigma$ errors for these parameters were obtained via a bootstrapping method that uses $10^3$ random half-sized subsamples of the 25 galaxies and determines the power-law parameters for every one of them. The standard deviations of the distributions for $\alpha_1$ and $\gamma_1$ are then divided by $\sqrt{2}$ to estimate 1-$\sigma$ confidence intervals for the full data set. The best power-law fit and its 1-$\sigma$ confidence interval are displayed in Fig.~\ref{fig_phenrel2}, while explicit numerical values are given in Section \ref{mapping}.

To first order, one would expect that $\ms$ depends linearly on $\mg$, if both masses scale linearly with the mass of the parent haloe. The over-proportional growth of $\ms$ ($\alpha_1=1.46\pm0.10>1$) could be explained by the fact that more massive galaxies are generally older and therefore could convert a larger fraction of hydrogen gas into stars.

\subsection{Scale radius versus stellar mass}\label{scalelengths}

\citet{Kregel2002} investigated a sample of 34 nearby edge-on spiral galaxies, drawn from the ESO-LV catalog \citep{Lauberts1989} using four selection criteria: (i) inclination $i\geq87\rm\,deg$, (ii) blue diameter $D_{25}^{\rm B}>2.2~\rm arcmin$, (iii) Hubble type from S0--Sd, (iv) only regular field galaxies, i.e.~no interacting systems, no warped or lopsided systems. This sample is complete in terms of sample selection \citep[see][]{Kregel2002,Davies1990}, but the sample volume is too small to contain rare objects. For each galaxy in the sample \citet{Kregel2002} determined the scale radius $\rd$ of the stellar disc from the I-band luminosity profiles. They also obtained the morphological Hubble type $T$ for each source from the Lyon/Meudon Extragalactic Database (LEDA). Additionally, we estimated the disc stellar masses $\msdisk$ from the I-band magnitudes of the disc components according to Eq.~(\ref{eqms}).

\begin{figure}
  \includegraphics[width=\columnwidth]{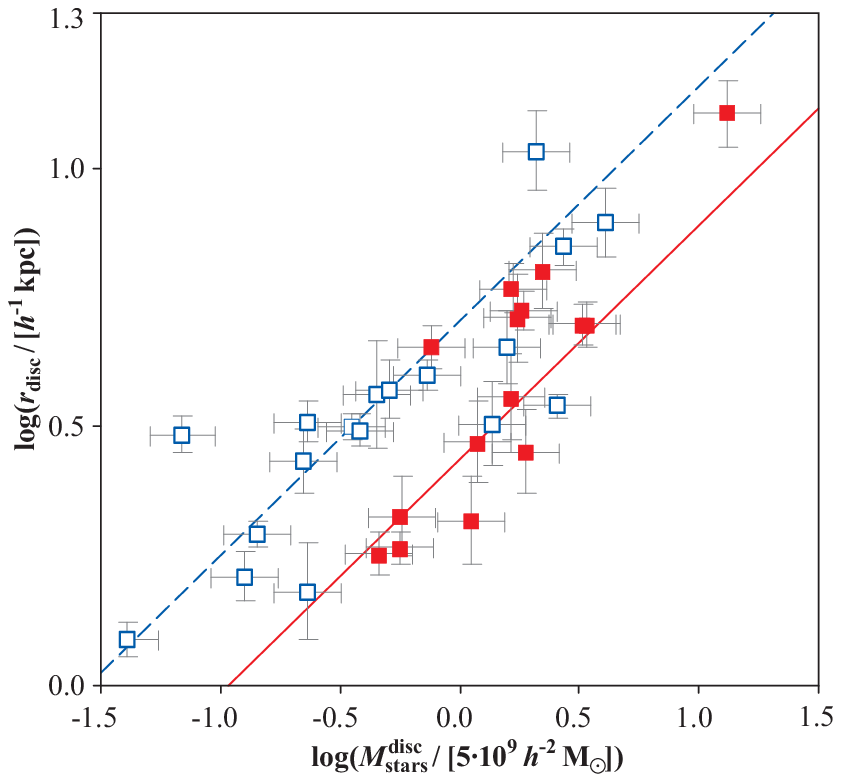}
  \caption{Relation between disc scale radius $\rd$ and disc stellar mass $\msdisk$. Squares represent 34 nearby spiral galaxies observed by \citet{Kregel2002}. Filled squares correspond to Sa/Sb-type galaxies, and empty squares represent Sc/Sd-type galaxies. These data were used to fit the free parameters of the model in Eq.~(\ref{phenrel2}). The solid line shows this model for $T=2$ (i.e.~Sab-type galaxies), while the dashed line shows the model for $T=6$ (i.e.~Scd-type galaxies). The slope of these power-laws is $\alpha_2=0.45\pm0.05$, consistent with a Freeman law \citep{McGaugh1995}.}
  \label{fig_phenrel2}
\end{figure}

Using these data, we investigated the relations between $\msdisk$, $\rd$ and $T$. The data points shown in Fig.~\ref{fig_phenrel2} suggest the approximate power-law with Hubble type correction of Eq.~(\ref{phenrel2}). The best fitting parameters $\alpha_2$, $\gamma_2$, and $\delta$ were obtained as in Section \ref{stvsgs} and explicit numerical values with errors are given in Section \ref{mapping}.

To first order, the $\rd$--$\msdisk$ relation can be understood in terms of a dark matter halo with an isothermal, singular, and spherical structure \citep[e.g.][]{Mo1998}. This model predicts that the virial radius $r_{\rm vir}$ is proportional to the cubic root of the dark matter mass $\mass_{\rm DM}$ at any fixed cosmic time. If $\rd$ were proportional to $r_{\rm vir}$ and $\msdisk$ were proportional to $\mass_{\rm DM}$, one would expect $\rd$ to scale as $(\msdisk)^{1/3}$. Our empirical result, $\alpha_2=0.45\pm0.05$, shows a slightly stronger scaling, consistent with the empirical Freeman law ($\alpha_2=0.5$), according to which disk galaxies have approximately constant surface brightness \citep{McGaugh1995}.

The secondary dependence of the $\rd$--$\msdisk$ relation on the Hubble type $T$ probably has multiple reasons: (i) early-type galaxies have more massive stellar bulges, which present an additional central potential that contracts the disc; (ii) bulges often form from disc instabilities, occurring preferably in systems with relatively low angular momentum, and hence early-type galaxies are biased towards smaller angular momenta and smaller scale radii; (iii) larger bulges, such as the ones of lenticular and elliptical galaxies, often arise from galaxy mergers, which tend to reduce the specific angular momenta and scale radii (see also \citealp{Obreschkow2009b}).

\label{lastpage}

\end{document}